\documentclass[english,aps,longbibliography,superscriptaddress,twocolumn,10pt,floatfix]{revtex4-1}
\usepackage[T1]{fontenc}
\usepackage[latin9]{inputenc}
\usepackage{geometry}
\usepackage{verbatim}
\usepackage{amsmath}
\usepackage{amssymb}
\usepackage{graphicx}
\usepackage[usenames]{color}
\usepackage{hyperref}

\makeatletter

\usepackage{babel}
\begin{document}

\title{Numerically exact full counting statistics of the nonequilibrium Anderson impurity model}

\author{Michael Ridley}

\affiliation{The Raymond and Beverley Sackler Center for Computational Molecular and Materials Science, Tel Aviv University, Tel Aviv 6997801, Israel}

\affiliation{School of Chemistry, Tel Aviv University, Tel Aviv 69978, Israel}

\author{Viveka N. Singh}

\affiliation{School of Chemistry, Tel Aviv University, Tel Aviv 69978, Israel}

\author{Emanuel Gull}

\affiliation{Department of Physics, University of Michigan, Ann Arbor, Michigan
48109, USA}

\author{Guy Cohen}

\affiliation{The Raymond and Beverley Sackler Center for Computational Molecular and Materials Science, Tel Aviv University, Tel Aviv 6997801, Israel}

\affiliation{School of Chemistry, Tel Aviv University, Tel Aviv 69978, Israel}
 \begin{abstract}
The time dependent full counting statistics of charge transport through an interacting quantum junction is evaluated from its generating function, controllably computed with the inchworm Monte Carlo method. Exact noninteracting results are reproduced; then, we continue to explore the effect of electron--electron interactions on the time-dependent charge cumulants, first-passage time distributions and $n$-electron transfer distributions. We observe a crossover in the noise from Coulomb blockade- to Kondo-dominated physics as the temperature is decreased. In addition, we uncover long-tailed spin distributions in the Kondo regime and analyze queuing behavior caused by correlations between single electron transfer events.
 \end{abstract}
\maketitle

\section{Introduction}
Mesoscopic quantum dots, molecules in junctions and other small quantum systems coupled to baths are often studied in transport experiments. It is possible to measure not only the electronic current through such systems, but also its noise and higher order moments. The complete set of moments determines the full counting statistics (FCS) of the system, which grants direct access to otherwise concealed properties: for example, the entanglement entropy, the fidelity, and thermoelectric efficiency fluctuations.

Apart from special limits, theoretical modeling of FCS in fermionic systems has so far been restricted to either noninteracting situations, or approximations whose accuracy is difficult to gauge. This difficulty is exacerbated in the experimentally relevant cases where strong electron--electron correlations and nonequilibrium effects are in play. A controlled and fully quantum description of the current and its moments is therefore sorely needed.

\begin{figure}[tbh]
\includegraphics{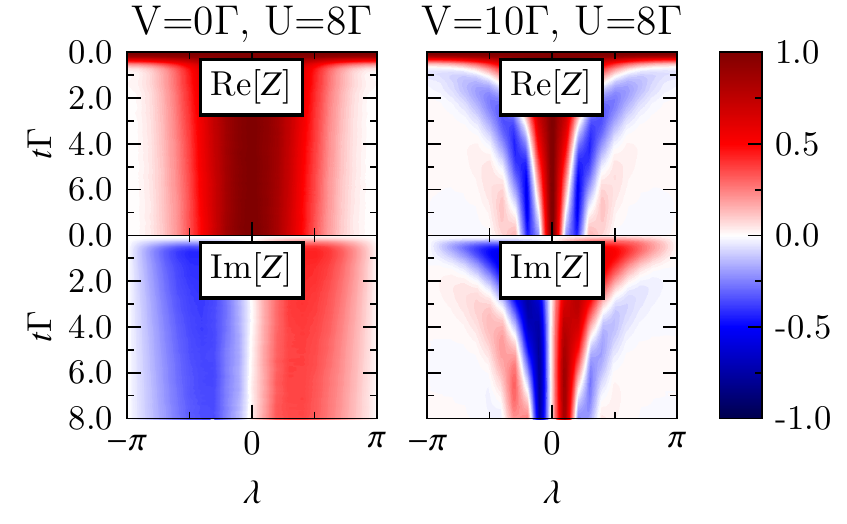}
\caption{\label{fig:ZContour_interacting} Real and imaginary parts of the generating function of lead population change $Z\left(\lambda,t\right)$ at inverse temperature $\beta=50/\Gamma$, for a strongly correlated quantum junction in the zero-bias (left) and finite-bias (right) cases.}
\end{figure}

In this paper, we provide such a description for a model of interacting fermions in a junction, within a numerically exact inchworm quantum Monte Carlo (iQMC) calculation. By computing the generating function of lead population change (Fig.~\ref{fig:ZContour_interacting}) we gain access to all moments of population transfer, first passage time distributions (FPTDs) and $n$-electron transfer distributions. We analyze Fano factors and consider the shapes of the distributions, which provide remarkably detailed information about the dynamics of single tunneling events. We extract indications of queuing effects in the presence of strong interactions.

The paper proceeds as follows. Section~\ref{sec:Overview} will motivate the full counting statistics, and Section~\ref{sec:FCS_theory} provide a brief overview of the main definitions and concepts used in FCS. In Section~\ref{sec:AIM} we describe how FCS can be obtained within iQMC, specializing to a particular model. Next, we present results for non-interacting (Sec.~\ref{sec:Results_nonint}) and interacting (Sec.~\ref{sec:Results_int}) systems. The effect of interactions, temperature and voltage on the first and second charge cumulants is discussed, and we further explore how first passage times and $n$-particle passage probabilities depend on the initial preparation. Finally, we discuss our conclusions in Section~\ref{sec:Conclusions}.

\section{Background and Motivation}\label{sec:Overview}

The transfer of charges across a small system, such as a molecular electronic junction or a mesoscopic quantum dot, is fundamentally a stochastic process \cite{Blanter2000}. Like any stochastic process, it is fully characterized by time dependent probability distributions \cite{vanKampen2007}; in this case, $P\left(\Delta n,t\right)$, where $\Delta n$ is the number of charges transferred by time $t$ \cite{Braggio2006}. Fluctuations in the electron current have long been known to provide experimentally accessible information not contained in the expectation value of the current \cite{Blanter2000,Landauer1998}, such as dwell times \cite{Beenakker2003} and transmission probabilities \cite{Vardimon2013}.

Current noise, the second moment of the current, has been of interest as a way to investigate reflection processes and traversal times in molecular junctions \cite{Selzer2013,Yang2014a,Ochoa2015,Ridley2017} and propagation of correlated electron--hole pairs in photon-assisted transport \cite{Rychkov2005,Battista2014,Zamoum2016}. A  great deal of work has focused on fluctuations in periodically driven systems, including the prediction \cite{Levitov1996,Dubois2013} and discovery \cite{Dubois2013a} of noiseless Lorentzian pulses, or ``levitons''. Quantum noise has also been used to extract effective quasiparticle charges $e^{*}$ of transmitted electrons due to Cooper pair formation in superconducting junctions ($e^{*}=2e$) \cite{Khlus1987,Lefloch2003a,Braggio2011}, the quantum Hall effect ($e^{*}=e/3$) \cite{Saminadayar1997}, or two-particle inelastic backscattering processes in the Kondo regime ($e^{*}=5e/3$) \cite{Sela2006,Golub2006,Golub2007,Zarchin2008,Yamauchi2011a,Ferrier2016}.

The noise-to-signal ratio, also known as the Fano factor $F$, can be used to characterize charge transfer statistics. It takes the value $F=1$ when the charge transfer is Poissonian (completely uncorrelated), as generally occurs for small transmission probabilities \cite{Levitov2004} and weak coupling between the molecule and bath \cite{Bomze2005}. Super-Poissonian shot noise ($F>1$) has been observed in single \cite{Safonov2003} and double \cite{Djuric2005,Kiesslich2007} quantum dot junctions and is generally associated with strong electron bunching and quasiparticle formation \cite{Blanter2000} that may occur in the transient \cite{Feng2008} or steady state \cite{Zarchin2008,Ferrier2016} regimes. In addition, the Fano factor can be tuned so that it is sub-Poissonian ($F<1$) in quantum dot \cite{Gustavsson2006,Emary2007} and graphene nanoribbon \cite{Tworzydlo2006} structures.

Instead of focusing only on the second moment of the current, it is possible to consider the full counting statistics (FCS) associated with the individual charge transfer events from an underlying moment-generating or characteristic function \cite{vanKampen2007}, and recover  \emph{all} higher-order moments and cumulants. This idea was pioneered by Levitov and Lesovik \cite{Levitov1993,Levitov1996}, who evaluated the steady state FCS for noninteracting systems. 

The FCS formalism for charge can be connected to entanglement entropy \cite{Klich2009} and fidelity \cite{Lesovik2006} in noninteracting systems, and was also generalized to a heat and work FCS giving access to efficiency fluctuations in thermoelectric junctions \cite{Esposito2015, agarwalla_full_2015}. Parallel with the development of the field of spintronics, the FCS of spin currents has been developed in regular metallic lead junctions \cite{dilorenzo2004,dilorenzo2005,dilorenzo2006,Fransson2010} and magnetic tunnel junctions (MTJs) \cite{Tang2014,Tang2017spin,Tang2017spinb}, and can be applied, for example, to the detection of spin-singlet pairs in the electron counting statistics \cite{dilorenzo2005,Schmidt2007}.  In the steady state, the second cumulant can be used to compute the zero-frequency current power spectrum \cite{Esposito2009,Sakano2011}, and in recent years a theory of FCS has been proposed which also gives access to frequency-dependent noise spectra \cite{Emary2007,Flindt2008,Ubbelohde2012}. Theoretical \cite{Tobiska2005,Saito2008,Simine2012} and experimental \cite{Pekola2015} investigations have demonstrated that the generating function satisfies steady state fluctuation--dissipation relations equivalent to the condition for detailed balance, although recent work indicates violation of detailed balance when bound states are established with circular probability currents in the steady state \cite{Stegmann2017} and modification of detailed balance when the system is undergoing transient evolution \cite{Altland2010}. It has also been shown that for noninteracting electrons, the zeros of the generating function all lie on the negative real axis \cite{Abanov2008,Abanov2009,Utsumi2013}, so that the position of zeros in the complex plane can shed light on interaction effects in the transport \cite{Kambly2011}.

Recently, the time dependence of higher-order cumulants has become accessible to experiment \cite{Flindt2009a,Fricke2010a,Ubbelohde2012}, motivating the study of transient cumulant generating functions and related objects which reveal more information on dynamics than the current or particle number. One such quantity, which is closely related to the generating function, is the waiting time distribution (WTD). The WTD is the probability distribution characterizing the time separating successive charge detection events \cite{Albert2011a,Albert2012,Thomas2014}. Experimentally, WTDs can be used to study the effect of coherence \cite{Brandes2008} and complex internal molecular structures such as double quantum dot \cite{Wang2007,Welack2008,Schaller2009,Welack2009} or normal--superconducting junctions exhibiting Andreev reflection \cite{Rajabi2013,Droste2016,Albert2016}, as these effects can alter the traversal times for electrons crossing the system. At the level of practical quantum electronics, these dynamical effects determine the maximum operation rate for devices \cite{Elzerman2004}.

Despite this flurry of theoretical activity, computing the FCS for generic interacting fermionic systems has so far proven to be challenging. 
In noninteracting systems, there now exist robust approaches to the study of both steady state and transient FCS, including an appealing coherent state path integral nonequilibrium Green's function formalism (PI-NEGF) \cite{Tang2014,Tang2014a,Yu2016,Tang2017} which can be perturbatively extended to interacting cases \cite{Souto2015,Tang2017,Tang2017a}.
Exact results are available for steady state FCS and noise characteristics of the AIM in the Fermi liquid regime and at the Toulouse point \cite{Komnik2005,Gogolin2006,Gogolin2006a,Golub2006,Golub2007,Gogolin2007}. The Fermi liquid theory is, however, valid only for low voltages and temperatures \cite{Sela2006,Gogolin2006a,Mora2008,Mora2009,Sakano2011b,Oguri2013}. 
Several approximate approaches to FCS in interacting systems have been successfully employed in appropriate regimes: a noncomprehensive list includes various pertubative expansions \cite{Tobiska2005,Sakano2011,Sakano2011b,Sakano2012,Oguri2013,Dong2013,Esposito2015,Gogolin2006,Schmidt2007,Schmidt2009,Park2011,Maier2011} and quantum master equation approaches \cite{Bagrets2003,Brandes2008,Flindt2008,Flindt2010,Simine2012,Albert2012,Kaasbjerg2015,Benito2016,Stegmann2017,Luo2017,Kosov2017,Wang2017}.
Nonperturbative numerical approaches also exist, but for models which do not capture the full complexity of interacting fermionic transport. These include hierarchical equations of motion in the spin-boson model \cite{Cerrillo2016} at high temperatures, and the density matrix renormalization group for the interacting resonant level model at its self dual point and at zero temperature \cite{Carr2011,Schmitteckert2014,Carr2015}. Also, an equilibrium determinant QMC method was recently applied to cold atomic gases  \cite{Humeniuk2017}.

To date, no numerically exact method has accessed FCS in a generic nonequilibrium interacting fermionic model; nor is a method currently available which is (even in principle) suitable for arbitrary temperatures, bias voltages and interaction strengths. In fact, this is largely true even for the calculation of the second cumulant, or noise. In this paper, we provide such a numerically exact method for computing the FCS, based upon the recently developed inchworm diagrammatic quantum Monte Carlo (iQMC) method \cite{Cohen2015}.

\section{FCS: theory and general considerations\label{sec:FCS_theory}}

In a charge detection experiment, one studies a system in which some regions, labeled $\ell$, are called ``leads''. Lead $\ell$ is found to contain $n_{\ell}$$\left(t\right)$ electrons on any given measurement of its total population at time $t$. The change in charge $\Delta n_{\ell}=n_{\ell}\left(t\right)-n_{\ell}\left(t_{0}\right)$ on the lead is measured between some initial time $t_{0}$ and the measurement time $t>t_{0}$. The FCS formalism aims to construct a generating function $Z\left(\lambda,t\right)$ for the probability distribution characterizing $\Delta n_{\ell}$, determined by the associated operator $\Delta\hat{N}_{\ell}$ (we suppress lead index $\ell$ from now on). For this purpose, an auxiliary counting field $\lambda$ is introduced, such that
\begin{equation}
Z\left(\lambda,t\right)\equiv\left\langle e^{i\lambda\Delta\hat{N}}\right\rangle =\sum_{\Delta n\in\mathbb{Z}}P\left(\Delta n,t\right)e^{i\lambda\Delta n}.\label{eq:Z}
\end{equation}
Here, $P\left(\Delta n,t\right)$ is the probability that the measured number of electrons has changed by the integer $\Delta n$ after time $t$. The distributions $P\left(\Delta n,t\right)$ can be obtained from the generating function via
\begin{equation}
P\left(\Delta n,t\right)=\int_{-\pi}^{\pi}\mathrm{d}\lambda\thinspace Z(\lambda,t)e^{-i\lambda\Delta n},\label{eq:Pn}
\end{equation}
and all moments $M_{k}\left(t\right)$ and cumulants $C_{k}\left(t\right)$ can be extracted by taking successive derivatives with respect to $\lambda$:
\begin{align}
M_{k}\left(t\right) & \equiv\left\langle \left(\Delta\hat{N}\right)^{k}\right\rangle =\lim_{\lambda\rightarrow0}\frac{\partial^{k}Z\left(\lambda,t\right)}{\partial\left(i\lambda\right)^{k}},\\
C_{k}\left(t\right) & \equiv \lim_{\lambda\rightarrow0}\frac{\partial^{k}\ln Z\left(\lambda,t\right)}{\partial\left(i\lambda\right)^{k}}.
\end{align}
The $M_{k}$ and $C_{k}$ contain equivalent information; in particular we note that $C_{1}=M_{1}$ and $ C_{2}=M_{2}-M_{1}^{2}$ relate the first and second moments. Of particular interest \cite{Tang2014,Tang2014a} is the probability distribution for no charges passing across the junction, which defines the \textit{idle time probability} $\Pi\left(t\right)$:
\begin{equation}
\Pi\left(t\right)\equiv P(0,t)=\int_{-\pi}^{\pi}\mathrm{d}\lambda\thinspace Z\left(\lambda,t\right).\label{eq:idletime}
\end{equation}
This can be used to extract the \textit{first passage time distribution} $W\left(t\right)$:
\begin{equation}
W\left(t\right)=-\frac{\mathrm{d}\Pi\left(t\right)}{\mathrm{d}t},\label{eq:WTD}
\end{equation}
which describes the probability distribution of the time separating initialization of the counting experiment from the first detection of a charge transfer event \cite{Albert2012}.

The counting statistics of any operator $\hat{A}\left(t\right)$ can be evaluated by way of an effective Hamiltonian transformation originally formulated by Levitov and Lesovik \cite{Levitov1993}. We will primarily be interested in the special case $\hat{A}\left(t\right)=\hat{N}_{L}$. The generating function for the moments of a change $\Delta a$ in the value associated with $\Delta\hat{A}\left(t\right)$ is given by \cite{Esposito2009}
\begin{equation}
Z\left(\lambda,t\right)=\textrm{Tr}\left[\hat{\rho}_{0}\hat{U}_{-\frac{\lambda}{2}}^{\dagger}\left(t,t_{0}\right)\hat{U}_{\frac{\lambda}{2}}\left(t,t_{0}\right)\right],\label{eq:ZFCS}
\end{equation}
in terms of the modified, operator-specific propagator $\hat{U}_{\gamma}\left(t,t_{0}\right)$, which is defined by \cite{Esposito2009}
\begin{equation}
\hat{U}_{\gamma}\left(t,t_{0}\right)\equiv e^{i\gamma\hat{A}\left(t\right)}\hat{U}\left(t,t_{0}\right)e^{-i\gamma\hat{A}\left(0\right)}.\label{eq:modified prop}
\end{equation}

The modified propagator itself can be written as a time-ordered integral with respect to a modified Hamiltonian:
\begin{equation}
\hat{U}_{\gamma}\left(t,t_{0}\right)=\mathrm{T}\exp\left[-\frac{i}{\hbar}\int_{t_{0}}^{t}\mathrm{d}\tau\thinspace\hat{H}_{\gamma}\left(\tau\right)\right],
\end{equation}
where
\begin{equation}
\hat{H}_{\gamma}\left(t\right)=e^{i\gamma\hat{A}\left(t\right)}\hat{H}\left(t\right)e^{-i\gamma\hat{A}\left(t\right)}-\hbar\gamma\partial_{t}\hat{A}\left(t\right).\label{eq:modham}
\end{equation}
This satisfies the relation $i\hbar\partial_{t}\hat{U}_{\gamma}\left(t,t_{0}\right)=\hat{H}_{\gamma}\left(t\right)\hat{U}_{\gamma}\left(t,t_{0}\right)$.
We express the generating function of Eq.~\eqref{eq:ZFCS} as a time ordered integral over the full Keldysh contour $\mathcal{C}=\left(\mathcal{C}_{+},\mathcal{C}_{-}\right)$, composed of an ``upper'' branch $\mathcal{C}_{+}$ and a ``lower'' branch $\mathcal{C}_{-}$ \cite{Keldysh1965}. Using the contour representation, $Z\left(\lambda,t\right)$ is given by the compact form
\begin{equation}
Z\left(\lambda,t\right)=\textrm{Tr}\left\{ \hat{\rho}_{0}\hat{T}_{\mathcal{C}}\left[\exp\left(-\frac{i}{\hbar}\int_{\mathcal{C}}\mathrm{d}z\thinspace\hat{H}_{\gamma}\left(z\right)\right)\right]\right\} ,\label{eq:zlambt}
\end{equation}
where $\hat{T}_{\mathcal{C}}$ orders times later on the contour to the left.

The auxiliary field $\gamma$ takes a different value on each branch:
\begin{equation}
\begin{aligned}\gamma\left(z\right) & =\frac{\lambda}{2},\ z\in\textrm{\ensuremath{\mathcal{C}_{+},}}\\
\gamma\left(z\right) & =-\frac{\lambda}{2},\ z\in\textrm{\ensuremath{\mathcal{C}_{-}.}}
\end{aligned}
\end{equation}
When $\lambda=0$, it is immediately apparent that the generating function $Z\left(\lambda,t\right)=1$ for all $t$. In general, $Z\left(\lambda,t\right)$ is periodic in $\lambda$ with a periodicity of $2\pi$. Any value of $\lambda\neq2\pi j$ for $j\in\mathcal{Z}$ means that the Hamiltonian depends on the contour branch, such that time reversal symmetry is explicitly broken.

\section{Model and Monte Carlo algorithm\label{sec:AIM}}

\subsection{Model and observables}

We now specialize the discussion to the concrete case of the nonequilibrium Anderson impurity model (AIM) \cite{Anderson1961}. The model's Hamiltonian can be written as follows:
\begin{equation}
\hat{H}=\hat{H}_{D}+\hat{H}_{B}+\hat{H}_{V}.\label{eq:Anderson}
\end{equation}
$\hat{H}_{D}$ describes a small, interacting ``dot'' subsystem; $\hat{H}_{B}$ depicts a set of large, noninteracting lead subsystems; and $\hat{H}_{V}$ is a hybridization Hamiltonian coupling the dot and the leads. These three terms take the following form:
\begin{align}
\hat{H}_{D} & =\underset{\sigma}{\sum}\varepsilon_{\sigma}\hat{d}_{\sigma}^{\dagger}\hat{d}_{\sigma}+U\hat{d}_{\uparrow}^{\dagger}\hat{d}_{\uparrow}\hat{d}_{\downarrow}^{\dagger}\hat{d}_{\downarrow},\label{eq:Hdot}\\
\hat{H}_{B} & =\underset{k\sigma,\ell}{\sum}\varepsilon_{k\sigma,\ell}\hat{a}_{k\sigma,\ell}^{\dagger}\hat{a}_{k\sigma,\ell},\label{eq:Hbath}\\
\hat{H}_{V} & =\underset{k\sigma,\ell}{\sum}\left(V_{k\sigma,\ell}\hat{a}_{k\sigma,\ell}^{\dagger}\hat{d}_{\sigma}+\textrm{H.c.}\right).\label{eq:Hhyb}
\end{align}
Here, the $\hat{d}_{\sigma}$ denote dot annihilation operators labeled by a spin index $\sigma$, while the $\hat{a}_{k\sigma,\ell}$ operators signify lead annihilation operators labeled by a spatial index $k$, spin index $\sigma$ and lead index $\ell $. We assume two leads, denoted $\ell=L,R$. The $\varepsilon_{\sigma}$ represent dot level energies, and the charging energy $U$ determines the strength of Coulomb repulsion between electrons. In this paper, we set $\varepsilon_{\sigma}=-\frac{U}{2}$ (the particle--hole symmetric case, to which our method is not in any way limited). The hybridization parameters $V_{k\sigma,\ell}$ couple the lead states to the dot and the $\varepsilon_{k\sigma,\ell}$ enumerate the energies of lead states. The system is initially prepared in the state $\rho_{0}=\rho_{L}\otimes\rho_{D}\otimes\rho_{R}$, a decoupled equilibrium state of $\hat{H}_{0}\equiv\hat{H}_{D}+\hat{H}_{B}$ in which the chemical potential in the leads is given by $\mu_{L}=V/2$, $\mu_{R}=-V/2$. $\rho_{D}$ is chosen to be in one of the four eigenstates of $H_{D}$: the empty state $\left|0\right\rangle $; the singly occupied spin up and spin down down states, $\left|\sigma\right\rangle $, with $\sigma=\uparrow,\thinspace\downarrow$; and the doubly occupied dot state $\left|\uparrow\downarrow\right\rangle $. These states will be denoted by the label $\phi$ in what follows.

The $\varepsilon_{k\sigma,\ell}$ and $V_{k\sigma,\ell}$ are completely determined by the coupling density
\begin{equation}
\Gamma_{\ell}\left(\omega\right)=\pi\underset{k}{\sum}\left|V_{k,\ell}\right|^{2}\delta\left(\omega-\varepsilon_{k\sigma,\ell}\right).\label{eq:density}
\end{equation}
We choose this to be a flat band with a smooth cut-off \cite{Werner2009}:
\begin{equation}
\Gamma_{\ell}\left(\omega\right)=\frac{\Gamma_{\ell}}{\left(1+e^{\nu\left(\omega-\Omega_{c}\right)}\right)\left(1+e^{-\nu\left(\omega+\Omega_{c}\right)}\right)}.\label{eq:fermdensity}
\end{equation}
In what follows, we set $\Gamma_{\ell}=\frac{1}{2}$ such that $\Gamma\equiv\underset{\ell}{\sum}\Gamma_{\ell}=1$ determines our unit of energy. We take the leads' band cutoff $\Omega_{c}$ to be $10\Gamma$, and their edge width $\frac{1}{\nu}$ to be $0.1\Gamma$.

At the initial time $t_{0}=0$ of the simulation, a coupling quench is performed and the subsystems are suddenly coupled by the introduction of $H_{V}$ into the Hamiltonian. This kind of switch-on is often referred to as the \textit{partitioned} approach, and has been contrasted in the quantum transport literature to \textit{partition-free} approaches (such as a voltage quench) in which the dot--lead coupling is established before a bias is added to the lead energies \cite{Stefanucci2004}. We note in passing that a partition-free voltage quench has also been explored within inchworm QMC \cite{Antipov2017}, by augmenting the finite Keldysh branches with an imaginary time Matsubara branch, thus obtaining the full Konstantinov--Perel' contour \cite{Konstantinov1960}.

In the present context, we are interested in the FCS for this model. We therefore set $\hat{A}\left(t\right)\rightarrow\hat{N}_{L}$ in the modified Hamiltonian of Eq.~\eqref{eq:modham}, where $\hat{N}_{L}\equiv\sum_{k\sigma}\hat{a}_{k\sigma,L}^{\dagger}\hat{a}_{k\sigma,L}$ is the particle number operator in lead $L$. This counting function describes the moments of population changes (rather than currents), but also provides access to the mean time-dependent current flowing out of the lead, $I_{L}\left(t\right)=\left\langle \mathrm{d}\hat{N}_{L}\left(t\right)/\mathrm{d}t\right\rangle $, by way of a time derivative. The current noise and higher order cumulants can only be accessed at steady state. For example, the noise
\begin{equation}
S_{LL}\left(\omega=0\right)\equiv\underset{t\rightarrow\infty}{\lim}\int\mathrm{d}\tau\thinspace\left\langle \Delta\hat{I}_{L}\left(t+\tau\right)\Delta\hat{I}_{L}\left(t\right)\right\rangle ,
\end{equation}
where $\Delta I_{L}=I_{L}-\left\langle I_{L}\right\rangle $, is related to the long time limit of the second cumulant \cite{Esposito2009}. We can therefore write:
\begin{align}
\underset{t\rightarrow\infty}{\lim}I_{L}\left(t\right) & =\underset{t\rightarrow\infty}{\lim}\frac{C_{1}\left(t\right)}{t},\label{eq:asympC1}\\
S_{LL}\left(\omega=0\right) & =\underset{t\rightarrow\infty}{\lim}\frac{C_{2}\left(t\right)}{t}.\label{eq:asympC2}
\end{align}
From the first and second cumulants, one can define the time-dependent Fano factor for the \textit{population} \cite{Kambly2013}
\begin{equation}
F\left(t\right)=\frac{C_{2}\left(t\right)}{C_{1}\left(t\right)}.\label{eq:Fanot}
\end{equation}
In the long-time limit, this converges to the Fano factor (up to a scaling factor of $2q$) \cite{Beenakker2003,Emary2007} for the \textit{current},
\begin{equation}
\frac{S_{LL}}{2qI_{L}}=\frac{S_{LL}}{S_{P}}=\underset{t\rightarrow\infty}{\lim}\frac{F\left(t\right)}{2q}.
\end{equation}
This is the Fano factor usually observed in steady state transport experiments, where $q$ is the charge of an individual carrier and $S_{P}=2qI_{L}$ is the Poisson value of the shot noise \cite{Blanter2000,Safonov2003,Beenakker2003}.

\subsection{FCS with inchworm Monte Carlo}

Within the modified Hamiltonian picture of Sec.~\ref{sec:FCS_theory}, the problem of evaluating a generating function can be mapped onto that of evaluating the time dependence of a unit operator propagated by the modified Hamiltonian. For the AIM with $\hat{N}_{L}$ as the counting field, using the Baker--Hausdorff lemma and the fact that $\hat{N}_{L}$ commutes with $\hat{H}_{0}$ one can show \cite{Tang2014} that the modified Hamiltonian of Eq.~\eqref{eq:modham} is equivalent to the AIM, but with lead--molecule coupling terms that acquire a branch-dependent phase:
\begin{equation}
\begin{aligned}\hat{H}_{\gamma}\left(z\right) & =e^{i\gamma\left(z\right)\hat{N}_{L}}\hat{H}e^{-i\gamma\left(z\right)\hat{N}_{L}}\\
 & =\hat{H}\left(\left\{ V_{k\sigma,\ell}\left(z;\lambda\right)\right\} \right),
\end{aligned}
\label{eq:modham-1}
\end{equation}
\begin{equation}
V_{k\sigma,\ell}\left(z;\lambda\right)=\begin{cases}
\begin{array}{c}
V_{k\sigma,\ell}e^{i\frac{\lambda}{2}\delta_{\ell L}},\ z\in\textrm{\ensuremath{\mathcal{C}_{+}},}\\
V_{k\sigma,\ell}e^{-i\frac{\lambda}{2}\delta_{\ell L}},\ z\in\textrm{\ensuremath{\mathcal{C}_{-}}}.
\end{array}\end{cases}\label{eq:modified coupling}
\end{equation}
The generating function in Eq.~\eqref{eq:zlambt} may then be evaluated by a standard Keldysh hybridization expansion \cite{Werner2006,Muhlbacher2008,Werner2009}. Whereas usually in such expansions an operator corresponding to some observable is applied at the fold of the Keldysh contour, here no such operator is needed, and the full counting statistics are given directly in terms of a modified propagator:
\begin{equation}
Z_{\phi}\left(\lambda,t\right)=p_{\phi\phi}\left(t^{+},t^{-};\lambda\right).\label{eq:Zpropagator}
\end{equation}
In this expression, $p_{\phi\phi'}\left(z_{1},z_{2};\lambda\right)$ is analogous to the dressed restricted propagators used in propagator noncrossing approximations \cite{Eckstein2010,Hartle2015,Chen2017I}, in bold-line QMC \cite{Gull2010,Gull2011,Cohen2013,Cohen2014,Cohen2014a},  and in inchworm QMC \cite{Cohen2015,Antipov2017,Chen2017I,Chen2017II,Dong2017},  but modified to include the auxiliary field via Eq.~\eqref{eq:modified coupling}:
\begin{equation}
\begin{aligned}p_{\phi\phi'}\left(z_{1},z_{2};\lambda\right) & \equiv\left\langle \phi\right|\textrm{Tr}_{B}\Biggl\{\rho_{0}e^{-i\int_{z_{2}}^{z_{1}}\hat{H}_{\lambda}\left(z\right)\mathrm{d}z}\Biggr\}\left|\phi'\right\rangle ,\\
\hat{H}_{\lambda}\left(z\right) & \equiv\hat{H}_{0}+\hat{H}_{V}\left(\left\{ V_{k\sigma,\ell}\left(z;\lambda\right)\right\} \right).
\end{aligned}
\label{eq:atomicpropagator}
\end{equation}

\begin{figure}
\includegraphics{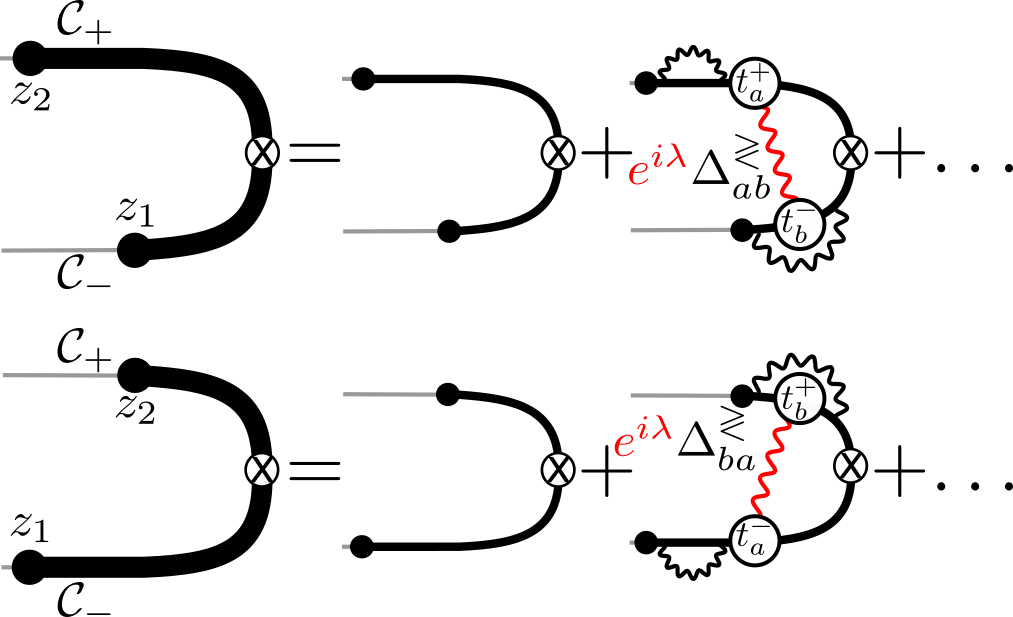}
\caption{\label{fig:inchworm_diagrams}Diagrammatic expansion of the full propagator (upper half of the panel) and the corresponding term in the propagator with contour branch indices exchanged. Thick lines denote a full propagator, thinner lines a bare propagator, wiggly lines a pair of hybridization events, and the ``X'' the fold in the Keldysh contour. The elements marked in red are those modified by the counting field, resulting in the breaking of time-reversal symmetry for $\lambda\neq0$.}
\end{figure}

When $p_{\phi\phi'}\left(z_{1},z_{2};\lambda\right)$ is expressed in the interaction picture and expanded in powers of the hybridization, it becomes a weighted sum over configurations $C$, defined by the set of contour times at which hybridization events (including dot and lead operators) occur. Each configuration sampled at a given diagram order $n$ corresponds to the insertion of $n$ hybridization lines connecting $2n$ ``kink'' times on the Keldysh contour where a particle of given spin $\sigma$ is added to or removed from the dot. We illustrate a diagram of order $n=0$ and another of order $n=3$ in Fig.~\ref{fig:inchworm_diagrams}, where the full propagator is represented by a thick black line, the bare atomic propagator by a thin black line and paired hybridization events by wiggly lines.

The full propagator can then be rewritten as a sum over diagrams, which can be stochastically sampled by diagrammatic Monte Carlo algorithms \cite{Gull2011b}:
\begin{equation}
p_{\phi\phi'}\left(z_{1},z_{2};\lambda\right)=\sum_{n=0}^{\infty}\sum_{C}w_{\mathrm{loc}}^{\left(n\right)}\left(C\right)w_{\mathrm{hyb}}^{\left(n\right)}\left(C;\lambda\right).\label{eq:propMC}
\end{equation}
Here the first summation is over the expansion order $n$ and the second summation is over all configurations $C=\left\{ z_{1},\ldots,z_{2n}\right\}$ corresponding to the $2n$ kink times at a given order. The $w_{\mathrm{loc}}$ are products of interacting (but purely local) atomic state propagators:
\begin{equation}
\begin{aligned}w_{\mathrm{loc}}^{\left(n\right)}\left(C\right) & =\left(-i\right)^{\left(n_{+}-n_{-}\right)}\\
 & \times\prod_{i=0}^{2n-1}p_{\phi_{i}\phi_{i+1}}^{\left(0\right)}\left(z_{i},z_{i+1}\right),\\
p_{\phi\phi'}^{\left(0\right)}\left(z_{1},z_{2}\right) & =\left\langle \phi\right|\textrm{Tr}_{B}\left\{ \rho_{0}e^{-i\int_{z_{1}}^{z_{2}}H_{D}\mathrm{d}z}\right\} \left|\phi'\right\rangle .
\end{aligned}
\end{equation}
Their sign depends on the number of kinks on each branch of the contour, $n^+$ and $n^- = 2n-n^+$. The $w_{\mathrm{hyb}}$ denote a determinant given by a sum over all possible lead operator pairings $X$. Each term in this sum is given by the product of hybridization functions $\tilde{\Delta}^{\lambda}=\sum_{\ell}\tilde{\Delta}_{\ell}^{\lambda}$ at the corresponding set of time pairs, with the sign given by the permutation generating that pairing \cite{Gull2011b}:
\begin{equation}
w_{\mathrm{hyb}}^{\left(n\right)}\left(C;\lambda\right)=\sum_{X}\mathrm{sign}\left(X\right)\prod_{i=0}^{n}\tilde{\Delta}^{\lambda}\left(z_{i},z_{X_{i}}\right).
\end{equation}
These hybridization functions, which are dressed by the counting field in a contour-time dependent manner, must be specified on the $z_{1}$ and $z_{2}$ axes:
\begin{equation}
\begin{aligned}\tilde{\Delta}_{\ell}^{\lambda}\left(z_{1},z_{2}\right) & =e^{-i\lambda\left(1-\delta_{\nu\nu^{\prime}}\right)\delta_{\ell L}}\theta_{\mathcal{C}}\left(z_{1}-z_{2}\right)\Delta_{\ell}^{>}\left(z_{1},z_{2}\right)\\
 & +e^{i\lambda\left(1-\delta_{\nu\nu^{\prime}}\right)\delta_{\ell L}}\theta_{\mathcal{C}}\left(z_{2}-z_{1}\right)\Delta_{\ell}^{<}\left(z_{1},z_{2}\right).
\end{aligned}
\label{eq:hybKeldysh-1}
\end{equation}
Here, $\theta_{\mathcal{C}}$ is the Heaviside function ordering times along the Keldysh contour; $\nu,\nu^{\prime}\in\left\{ \mathcal{C}_{+},\mathcal{C}_{-}\right\} $ are the branch indices of $z_{1}$ and $z_{2}$, respectively; and $L$ is the lead for which FCS is being evaluated. Due to the modified coupling in Eq.~\eqref{eq:modified coupling}, the phase factors in the dressed hybridization function Eq.~\eqref{eq:hybKeldysh-1} cancel when $z_{1}$ and $z_{2}$ lie on the same branch, such that only hybridization lines that cross the folding time $t_{\mathrm{max}}$ of the Keldysh contour are modified by the counting field (see the red elements in Fig.~\ref{fig:inchworm_diagrams}). The undressed lesser and greater hybridization components,
\begin{equation}
\begin{aligned}\Delta_{\ell}^{<}\left(t_{1},t_{2}\right) & =-i\int_{-\infty}^{\infty}\frac{\mathrm{d}\omega}{\pi}e^{-i\omega\left(t_{1}-t_{2}\right)}\\
 & \times\Gamma_{\ell}\left(\omega\right)f\left(\omega-\mu_{\ell}\right),\\
\Delta_{\ell}^{>}\left(t_{1},t_{2}\right) & =i\int_{-\infty}^{\infty}\frac{\mathrm{d}\omega}{\pi}e^{-i\omega\left(t_{1}-t_{2}\right)}\\
 & \times\Gamma_{\ell}\left(\omega\right)\left[1-f\left(\omega-\mu_{\ell}\right)\right],
\end{aligned}
\end{equation}
can be expressed in terms of the level width function $\Gamma_{\ell}\left(\omega\right)$ as defined in Eq.~\eqref{eq:density}, and are parametrized only by the physical times $t_{1}$ and $t_{2}$ corresponding to the contour times $z_{1}$ and $z_{2}$.

Exact numerical approaches to the investigation of dynamics in quantum many-body systems typically suffer from an exponential scaling of computational cost with time. In real time Monte Carlo methods, this manifests in the dynamical sign problem: an exponential growth of stochastic errors with increasing time. However, the inchworm algorithm, as applied to the real time hybridization expansion \cite{Werner2006,Muhlbacher2008,Werner2009,Schiro2010,Gull2011,Gull2011b}, has been shown to bypass the dynamical sign problem in a wide variety of parameters. This was not only shown for the AIM \cite{Cohen2015,Antipov2017}, but also in the spin--boson model \cite{Chen2017I,Chen2017II} and---within the dynamical mean field approximation---for lattice models \cite{Dong2017}.

\begin{figure}
\includegraphics[width=8.6cm]{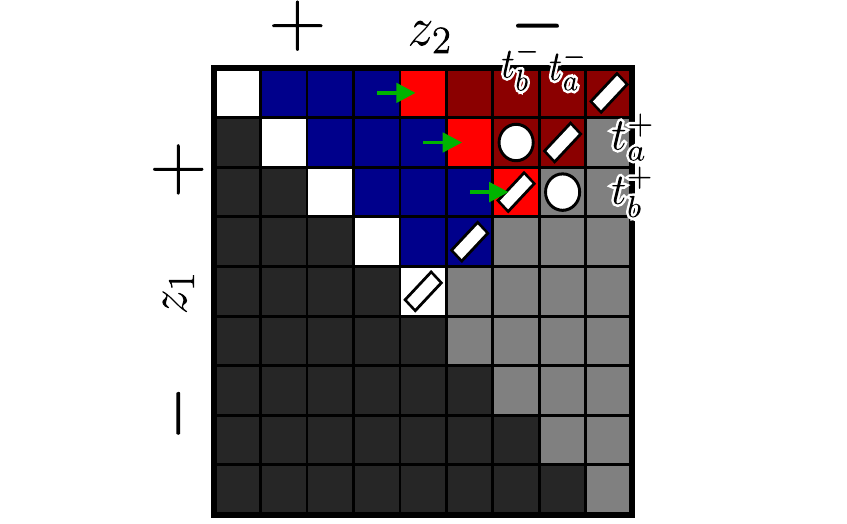}
\includegraphics[width=8.6cm]{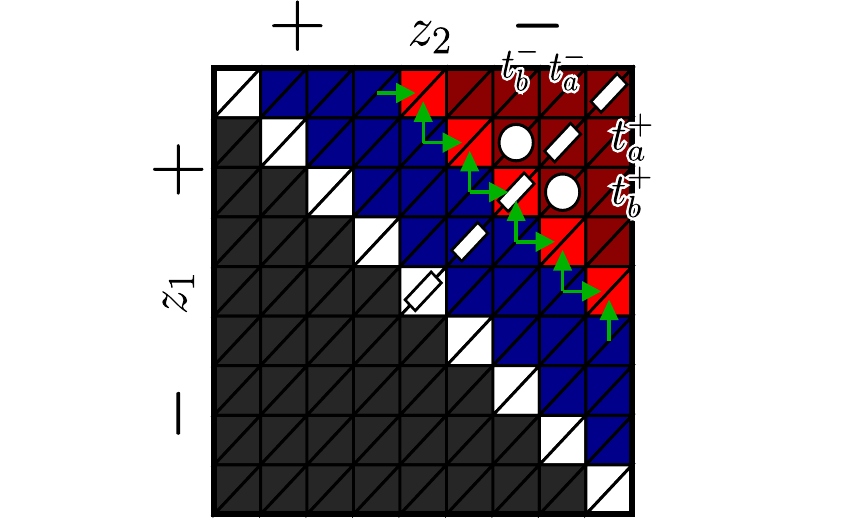}
\caption{\label{fig:inchworm_propagation}
Schematic illustration of an intermediate step in the inchworm in the inchworm propagation scheme when (top) time-reversal symmetry applies, and (bottom) when it is broken by the counting field. Here, each square denotes a propagator restricted to lie between two time arguments. Dark gray squares are unnecessary propagators, white squares are trivial propagators, blue squares are known propagators from previous steps, and red squares are unknown propagators. The lighter red squares are propagators which can be calculated in the present step, with the green arrows representing propagation direction. The dashed white line marks the symmetry element of time-reversal, and the white circles are two propagators connected by time reversal, as illustrated diagrammatically in Fig.~\ref{fig:inchworm_diagrams}.}
\end{figure}

Here, the inchworm algorithm is used to efficiently sum the expansion described above, by reusing data obtained from propagators on shorter time intervals to construct propagators on longer time intervals \cite{Cohen2015}. In the top panel of Fig.~\ref{fig:inchworm_propagation}, a snapshot of an intermediate step in the inchworm procedure is illustrated for a branch-independent Hamiltonian without a counting field. Each of the $z_{1}$ and $z_{2}$ axes shown consists of an ``unfolded'' Keldysh contour containing times ordered according to the structure $\left[t_{0}^{+},...,t_{max},...,t_{0}^{-}\right]$. In each inchworm step, propagators are extended along one time direction (this is called ``inching''), marked by the green arrows going to the right. In dark gray, we show time points unneeded for the calculation (as we can always select one contour time to appear later on the contour than the other). The white squares correspond to equal time propagators, which can be evaluated analytically. The blue squares correspond to time arguments for which the propagator is already known from previous steps, while the light red squares correspond to arguments at which the propagator may be evaluated immediately, given the currently known propagators. The dark red squares represent values to be computed at a later steps. In light gray, we show time points which can be obtained from reflection about the $z_{1}=-z_{2}$ axis (marked by a white dashed diagonal line), which can be seen as \emph{time reversal symmetry}: the propagators on one side of this symmetry element can be obtained from those on the other via the relation
\begin{equation}
p_{\phi\phi'}\left(z_{1},z_{2}\right)=p_{\phi'\phi}^{*}\left(\tilde{z}_{2},\tilde{z}_{1}\right).\label{eq:symmetryTR}
\end{equation}
Here, a contour time argument marked by a tilde, $\tilde{z}$, denotes the same physical time as that of $z$, but on the opposite contour.

Due to time reversal symmetry, in the original formulation of the inchworm algorithm the total number of propagators computed is $\frac{1}{4}\left(\frac{t}{\varDelta t}\right)^{2}$, as only one quadrant of the two time plane needs to be explicitly evaluated. Extending a propagator to the right in the quadrant used is identical to extending a symmetrically placed propagator up in the mirrored quadrant, so that time reversal symmetry is also maintained in the inching direction.

In diagrammatic terms, time reversal symmetry is illustrated by in Fig.~\ref{fig:inchworm_diagrams}. In the upper part of \ref{fig:inchworm_diagrams}, the cross-branch hybridization line corresponds to a factor of $\Delta^{>}\left(t_{b},t_{a}\right)$ in the propagator, which depends only on physical times and not on the branch indices. When the contour branches of all times appearing in this diagram are flipped, one obtains the expansion in the lower part of Fig.~\ref{fig:inchworm_diagrams}. For example, the hybridization factor $\Delta^{>}\left(t_{a},t_{b}\right)$ is replaced by $\Delta^{>}\left(t_{b},t_{a}\right)=-\left[\Delta^{>}\left(t_{a},t_{b}\right)\right]^{*}$. All other factors are similarly conjugated, and so Eq.~\eqref{eq:symmetryTR} holds.

By contrast, in the presence of the counting field we evaluate the propagator for nonzero $\lambda$, and the two contour-reflected diagrams are modified as illustrated by the red elements in Fig.~\ref{fig:inchworm_diagrams}: the dressed cross-branch hybridization factor, denoted by the red wiggly line, is replaced by $e^{i\lambda}\Delta^{>}\left(t_{a},t_{b}\right)$ (upper diagram) and $e^{i\lambda}\Delta^{>}\left(t_{b},t_{a}\right)=-e^{i\lambda}\left[\Delta^{>}\left(t_{a},t_{b}\right)\right]^{*}$ (lower diagram). Due to the unconjugated prefactor $e^{i\lambda}$, Eq.~\eqref{eq:symmetryTR} is no longer valid, and the two quadrants to either side of the $z_{1}=-z_{2}$ line must be evaluated separately, as illustrated in the lower panel of Fig.~\ref{fig:inchworm_propagation}. Since we no longer mirror the data or enforce the symmetry, an additional spurious numerical breaking of time reversal symmetry due to the asymmetry in inching direction can occur. To avoid this, we inch to each time point from the two possible directions and set the result to the average (illustrated by the green arrows, now going both right and up). This increases the number of simulations by an additional factor of two, so that $\left(\frac{t}{\varDelta t}\right)^{2}$ are needed in total, but the overall quadratic scaling with the time step remains. We have found that bidirectional inching generally provides higher quality data for a given amount of computer time, even without the counting field.

\section{Benchmark and validation for the non-interacting system}\label{sec:Results_nonint}
\subsection{Noninteracting FCS}

In the absence of interactions (when $U=0$), the FCS following the coupling quench can be exactly obtained by the path integral nonequilibrium Green's function (PI-NEGF) method, which provides an exact solution in the absence of electron--electron interactions \cite{Tang2014}. The generating function for a noninteracting AIM is given by the ratio of two Fredholm determinants:

\begin{equation}
Z\left(\lambda,t\right)=\det\left(\tilde{G}^{-1}\left(z_{1},z_{2}\right)\right)/\det\left(G^{-1}\left(z_{1},z_{2}\right)\right).\label{eq:zpinegf}
\end{equation}
Here, $G$ is the matrix of two-time Green's functions for the full molecular junction in Keldysh-rotated space \cite{Keldysh1965,Kamenev2011}, such that the times $z_{1},z_{2}$ correspond to pairs of times on the Keldysh contour with a folding point at $t_{\mathrm{max}}=t$. The Green's function is given by
\begin{equation}
G^{-1}\left(z_{1},z_{2}\right)=g^{-1}\left(z_{1},z_{2}\right)-\Delta_{L}\left(z_{1},z_{2}\right)-\Delta_{R}\left(z_{1},z_{2}\right),\label{eq:Ginv}
\end{equation}
where $g$ denotes the Green's function of the isolated dot and $\tilde{G}=G-\left(\tilde{\Delta}_{L}-\Delta_{L}\right)$ is a modified Green's function that depends upon the counting field $\lambda$ in the left lead. When expressed in the Keldysh-rotated representation, the two hybridizations are related via
\begin{equation}
\tilde{\Delta}_{\ell}\left(z_{1},z_{2}\right)=\exp\left(i\sigma_{x}\frac{\lambda}{2}\right)\Delta_{\ell}\left(z_{1},z_{2}\right)\exp\left(-i\sigma_{x}\frac{\lambda}{2}\right),\label{eq:lambdahyb}
\end{equation}
where $\sigma_{x}$ is the Pauli spin matrix. It is easy to show that this is equivalent to Eq.~\eqref{eq:hybKeldysh-1} when the Keldysh rotation is reversed. Further details on the implementation of Eq.~\eqref{eq:zpinegf} can be found in Refs.~\onlinecite{Tang2014,Tang2014a,Yu2016}. It was also shown in these works that the expression (\ref{eq:zpinegf}) correctly reduces to the Levitov--Lesovik formula for the generating function in the long-time limit \cite{Levitov1993,Levitov1996}, which can be written as
\begin{align}
Z\left(\lambda,t\right) & =\exp\left\{ t\int\frac{\mathrm{d}\omega}{2\pi}~\overset{\infty}{\underset{k=1}{\sum}}\frac{\left(-1\right)^{k+1}}{k}\textrm{Tr}\left[T\left(\omega\right)\right.\right.\label{eq:levlesexpand-1}\\
 & \times\left[\left(e^{i\lambda}-1\right)\left(1-f_{R}\left(\omega\right)\right)f_{L}\left(\omega\right)\right.\nonumber \\
 & \left.\left.\left.+\left(e^{-i\lambda}-1\right)\left(1-f_{L}\left(\omega\right)\right)f_{R}\left(\omega\right)\right]\right]^{k}\right\} .\nonumber 
\end{align}
Here, $f_{\ell}$ denotes the Fermi function of lead $\ell$ and $T\left(\omega\right)$ is the transmission probability for electrons to pass through the molecule.

In this work, the PI-NEGF generating function is computed for the noninteracting case without spin degeneracy, as in Ref.~\onlinecite{Tang2014}. However, we are interested in the AIM. The $U=0$ generating function for the AIM can be constructed from that of the single-electron model as a product:
\begin{equation}
Z_{\sigma,\sigma'}^{\left(AIM\right)}\left(\lambda,t\right)=Z_{\sigma}^{\left(dot\right)}\left(\lambda,t\right)Z_{\sigma'}^{\left(dot\right)}\left(\lambda,t\right).\label{eq:Zdotproduct}
\end{equation}
The $\sigma$, $\sigma^{\prime}$ subscripts denote the two spins. In the cases to be shown here, the system is initially decoupled from the leads with the dot in the unoccupied state $\left|0\right\rangle $, such that in the noninteracting calculations, each independent model begins with an unoccupied dot. The dot--bath coupling is then turned on at the quench time $t_{0}=0$.

\subsection{QMC results for the non-interacting system}

\begin{figure}
\includegraphics{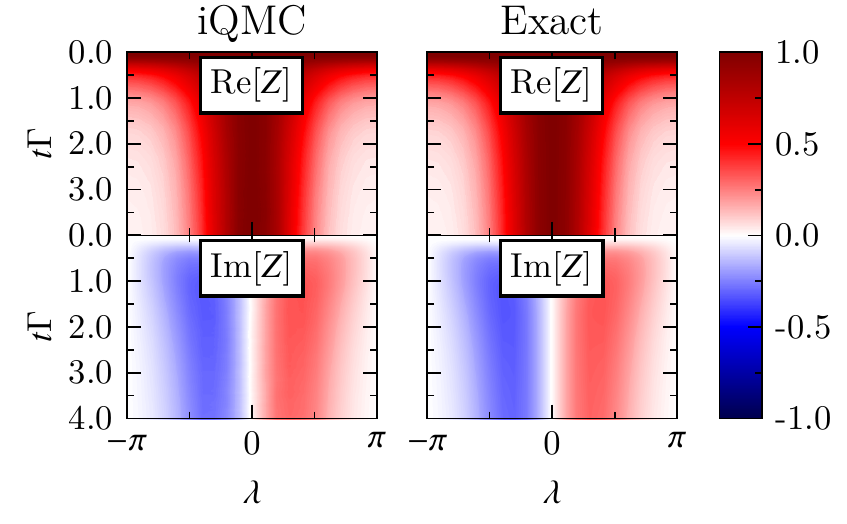}\caption{\label{fig:V0U0_contour}Real and imaginary parts of the generating function from iQMC (left) compared to the exact result (right) at $U=0\Gamma$, $V=0\Gamma$ and $\beta=50/\Gamma$.}
\end{figure}

\begin{figure}
\includegraphics{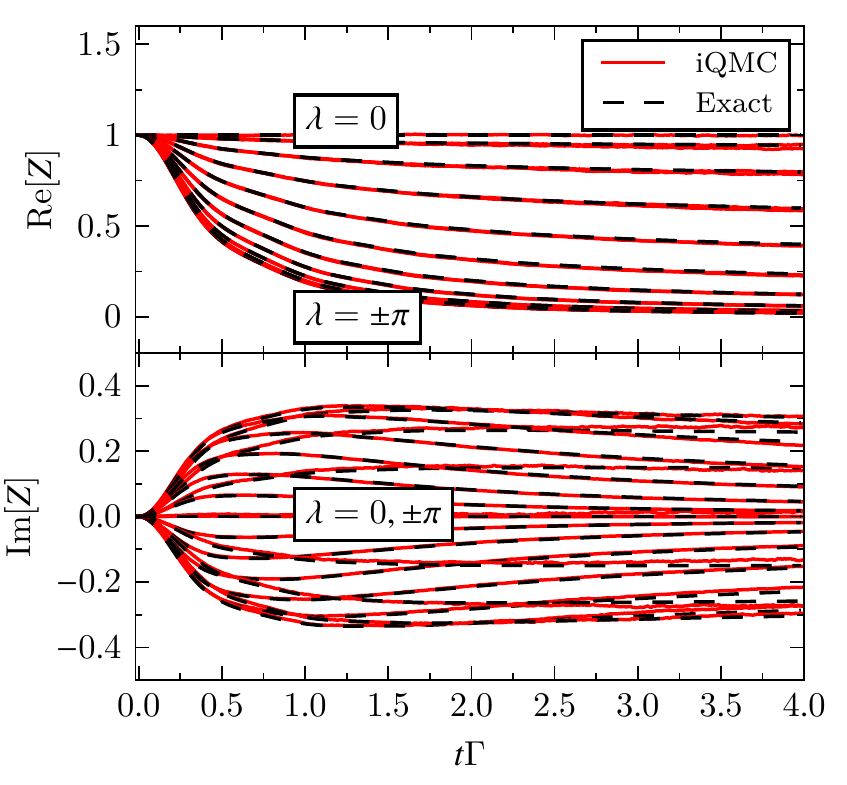}\caption{\label{fig:V0U0_cuts}Real and imaginary parts of the generating function from iQMC (solid red) compared to the exact result (dashed black) at $U=0\Gamma$, $V=0$ and $\beta=50/\Gamma$.}
\end{figure}

To verify the correctness of the iQMC method, we begin by performing calculations for the noninteracting case $U=0$ and comparing the results with exact results obtained from the PI-NEGF method. This formalism provides us with an independent verification of the new method: as the iQMC calculation is based on a hybridization (rather than an interaction) expansion, the noninteracting case is nontrivial and embodies a rigorous test of the method's accuracy \cite{Werner2006}. In Fig.~\ref{fig:V0U0_contour} we present the time evolution of $Z\left(\lambda,t\right)$. The two panels on the left are generated from iQMC, while the two panels on the right are exact results. As might be expected, $Z\left(\lambda=0,t\right)=1$ is satisfied to within the numerical errors. The following symmetry relations are also apparent:
\begin{equation}
\begin{aligned}\textrm{Re}\left[Z\left(\lambda,t\right)\right] & =\textrm{Re}\left[Z\left(\pi-\lambda,t\right)\right],\\
\textrm{Im}\left[Z\left(\lambda,t\right)\right] & =-\textrm{Im}\left[Z\left(\pi-\lambda,t\right)\right].
\end{aligned}
\label{eq:Zsym}
\end{equation}
We note that the time evolution of both real and imaginary parts of $Z$ exhibit monotonic decay towards zero after a transient timescale on the order of $1/\Gamma$.

In order to explore the correspondence between the numerical data and the exact result in more detail, we plot a series of cuts across the data in Fig~\ref{fig:V0U0_cuts}. Each curve shown here corresponds to a different value of $\lambda$ in the interval $\left[-\pi,\pi\right]$. The results from iQMC are shown in solid red, together with exact data in dashed black. The two sets of curves appear to be in very good agreement, with slight numerical deviations (on the order of 1\%) in the iQMC results visible at long times.

\begin{figure}
\includegraphics{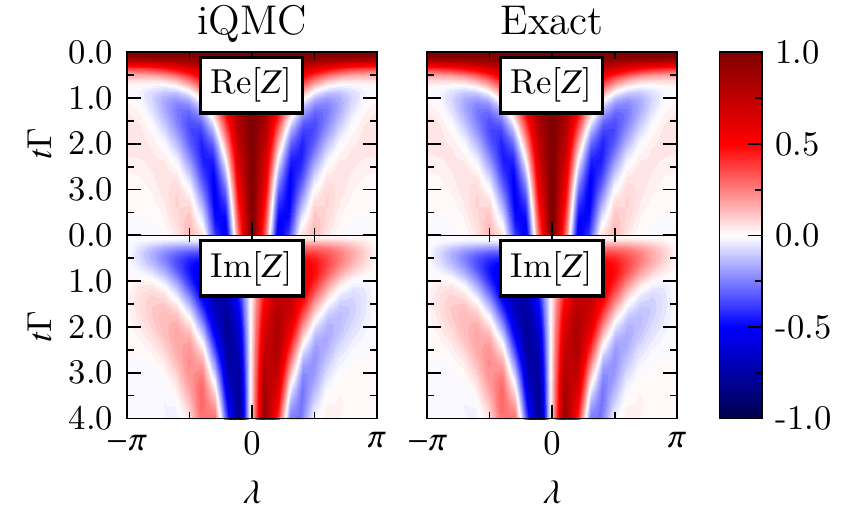}
\caption{\label{fig:U0V5_contour}Real and imaginary parts of the generating function from iQMC (left) compared to the exact result (right) at $U=0\Gamma$, $V=10\Gamma$ and $\beta=50/\Gamma$.}
\end{figure}

\begin{figure}
\includegraphics{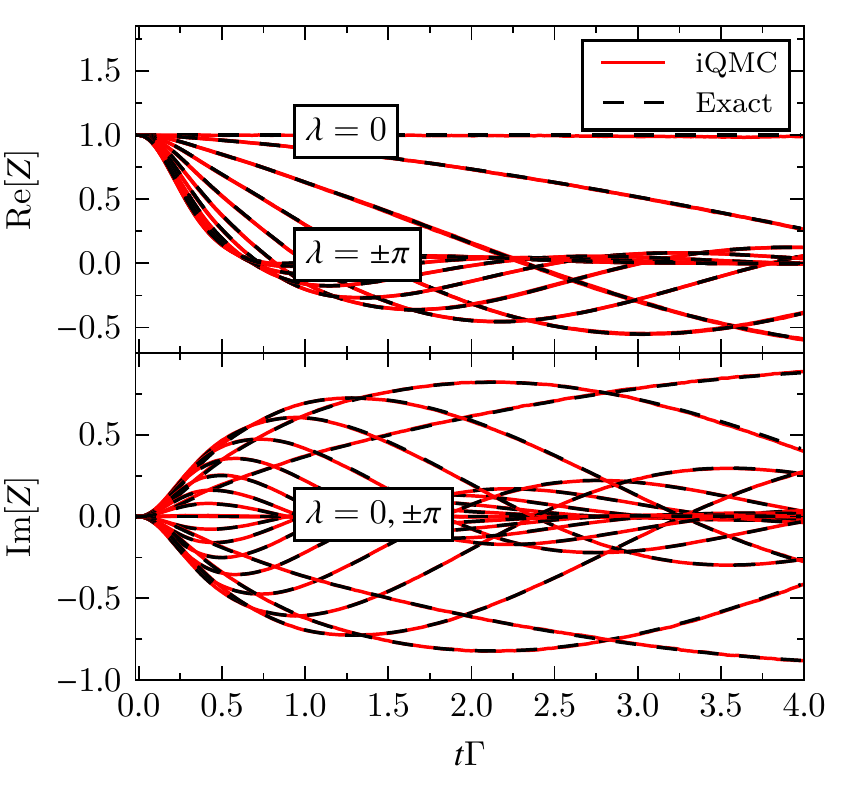}\caption{\label{fig:U0V5_cuts}Real and imaginary parts of the generating function from iQMC (solid red) compared to the exact result (dashed black) at $U=0\Gamma$, $V=10\Gamma$ and $\beta=50/\Gamma$.}
\end{figure}

The introduction of a finite voltage causes the components of the generating function to strongly oscillate as a function of time, as shown in Fig.~\ref{fig:U0V5_contour}. This indicates oscillations in cumulants of all orders, reflecting a universal phenomenon predicted by Flindt \textit{et al.} in Ref.~\onlinecite{Flindt2009a} and later observed experimentally \cite{Fricke2010a}. Physically, the introduction of a finite bias enhances the short-time ``ringing'' behavior, as has previously been observed for the lowest cumulant (the current) \cite{Hartle2015,Ridley2015}. As the FCS in the presence of a voltage bias is substantially richer, it is instructive to consider a detailed comparison of specific cuts again (see Fig.~\ref{fig:U0V5_cuts}). Remarkably, it is observable from this figure (which used less computational resources than Fig~\ref{fig:V0U0_cuts}) that at higher voltages it is easier to converge the numerics. This is a notable property of real time iQMC \cite{Cohen2011,Cohen2013}, which makes its regime of efficient applicability different from that of, \emph{e.g.}, low-energy wavefunction methods.


Knowledge of the generating functional provides access to all moments and cumulants in principle, but the numerical exercise of obtaining them in practice may not be trivial. One of the main technical difficulties stems from the fact that within iQMC, the relative error in this data is generally larger at small $\lambda$ values. This is for two reasons: first, part of the statistical iQMC error is independent of the value of the observable. This source of error is strongly mitigated by the inchworm algorithm, but not completely removed, and results in noisy numerical derivatives. Second, there is a small systematic error due to the time discretization. This error is difficult to fully converge, and can sometimes be significant: for example, consider how the imaginary part of $Z\left(0,t\right)$ visibly deviates from its exact value of zero in the lower panel of Fig.~\ref{fig:V0U0_cuts}.

Therefore, if we evaluate $\lambda$ derivatives of $Z\left(\lambda,t\right)$ in the limit of small $\lambda$ by a finite difference formula $\lim_{\lambda\rightarrow0}\frac{\mathrm{d}Z\left(\lambda,t\right)}{\mathrm{d}\lambda}\simeq\frac{Z\left(\Delta\lambda,t\right)-Z\left(0,t\right)}{\Delta\lambda}$, the value of $\Delta\lambda$ cannot easily be taken to zero without investing unjustified amounts of computer time. Furthermore, the oscillatory form of the function hints that an increasingly small value will be needed as one propagates to longer times.

However, the dependence of the moments and cumulants on $\Delta\lambda$ can be understood at long times by considering the Levitov--Lesovik formula for the current cumulant generating function $S\left(\lambda\right)\equiv\underset{t\rightarrow\infty}{\lim}\frac{\ln Z\left(\lambda,t\right)}{t}$, with $Z\left(\lambda,t\right)$ given by Eq.~\eqref{eq:levlesexpand-1}. If this is inserted into the finite difference derivative with respect to a small shift $\Delta\lambda$, and the integrand is expanded to linear order in $\Delta\lambda$, one arrives at the following expressions for the logarithmic and direct derivatives:
\begin{align}
\left.\frac{d\ln Z\left(\lambda,t\right)}{d\left(i\lambda\right)}\right|_{\lambda=0} & =\underset{\Delta\lambda\rightarrow0}{\lim}t\int\frac{\mathrm{d}\omega}{2\pi}\\
 & \times\textrm{Tr}\left[T\left(\omega\right)\left(f_{L}\left(\omega\right)-f_{R}\left(\omega\right)\right)\right],\nonumber \\
\left.\frac{dZ\left(\lambda,t\right)}{d\left(i\lambda\right)}\right|_{\lambda=0} & =\underset{\Delta\lambda\rightarrow0}{\lim}\frac{1}{i\Delta\lambda}\\
 & \times\left[\exp\left\{ i\Delta\lambda t\int\frac{d\omega}{2\pi}\textrm{Tr}\left[T\left(\omega\right)\right.\right.\right.\nonumber \\
 & \times\left.\left.\left.\left(f_{L}\left(\omega\right)-f_{R}\left(\omega\right)\right)\right]\right\} -1\right].\nonumber 
\end{align}
Thus, for finite but small $\Delta\lambda$ the first cumulant (which corresponds to Eq.~\eqref{eq:asympC1}) exhibits no dependence on $\Delta\lambda$, whereas the first moment oscillates at long times as $\exp\left(\sim i\Delta\lambda t\right)$. This implies that numerical derivatives taken from $\ln Z$ will converge at a finite $\Delta\lambda$ in the long time limit, while the error in numerical derivatives taken directly from the generating function will diverge.

\begin{figure}
\includegraphics{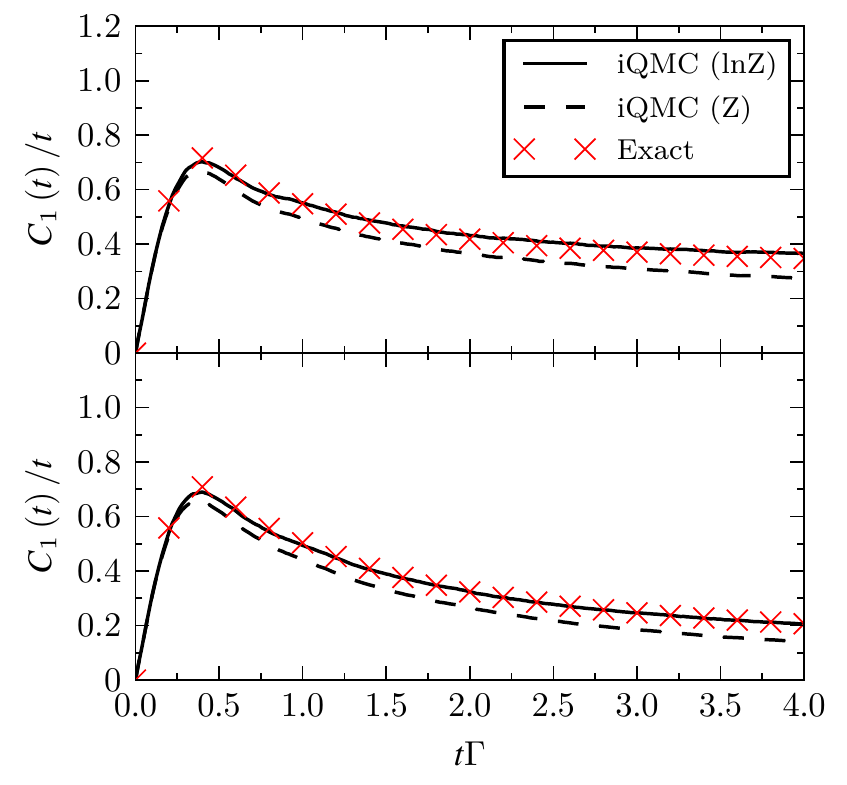}\caption{\label{First-moments}First cumulants of the particle number in the noninteracting case $U=0$ are shown, as obtained from iQMC. Finite difference derivatives of $\ln Z$ (thick lines) and of $Z$ (dashed lines) are plotted against the essentially exact benchmark PI-NEGF logarithmic derivative for $\Delta\lambda$=0.001 (red crosses). The inverse temperatures are $\beta=0.4/\Gamma$ (top) and $\beta=50/\Gamma$ (bottom).}
\end{figure}

To see this in practice, consider Fig.~\ref{First-moments}. Here, dot populations normalized by time ($C_{1}\left(t\right)/t$) are plotted. These plots were obtained from both the logarithmic (thick black lines) and direct (dashed black lines) derivatives with a symmetric bias of $V=1\Gamma$. Cumulants are obtained from both derivatives for $\Delta\lambda=0.6$ and compared with the exact result (red crosses) obtained from PI-NEGF at $\Delta\lambda=0.001$. This is done for high temperatures ($\beta=0.4\Gamma$, upper panel) and low temperatures ($\beta=50\Gamma$, lower panel). As suggested by the preceding analysis, the convergence to the $\Delta\lambda\rightarrow0$ limit is substantially faster when logarithmic derivatives are taken. While the direct and logarithmic derivatives are in agreement at short times, the direct derivative diverges from the exact result at long times. Without performing a full error analysis, it is difficult to determine whether the logarithmic derivative differs significantly from the exact result. At low temperatures convergence is slower, and the steady state value (equivalent to the long time limit of the left current) increases.

\begin{figure}
\includegraphics{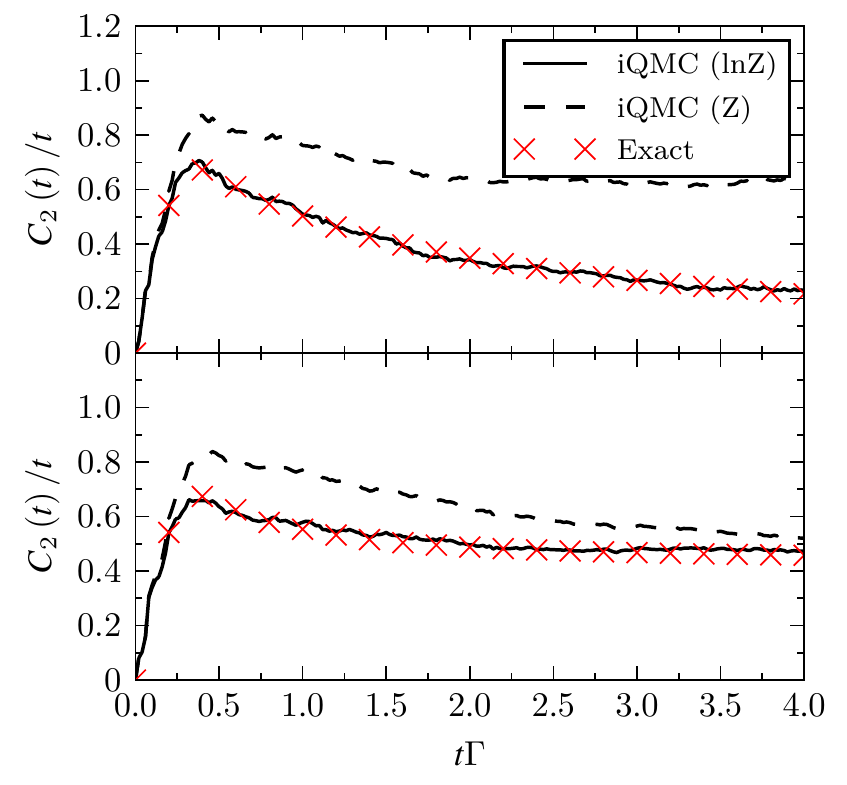}\caption{\label{Secondmoments}Second cumulants of the particle number in the noninteracting case $U=0\Gamma$ are shown, as obtained from iQMC. Cumulants obtained from finite difference derivatives of $\ln Z$ (thick lines) and of $Z$ (dashed lines) are plotted against the essentially exact benchmark PI-NEGF logarithmic derivative for $\Delta\lambda$=0.001 (red crosses). The inverse temperatures are $\beta=0.4/\Gamma$ (top) and $\beta=50/\Gamma$ (bottom).}
\end{figure}

We repeat this analysis for the normalized second cumulants $C_{2}\left(t\right)/t$ of the population in Fig.~\ref{Secondmoments}. The second cumulant $C_{2}\left(t\right)$ increases linearly with time \cite{Flindt2009a,Esposito2009}, and is related at long times to the current noise in the left lead $S_{LL}$ via Eq.~\eqref{eq:asympC2}. The cumulant extracted from the iQMC data is in excellent agreement with the exact result event for relatively large values of $\Delta \lambda$ when the cumulant is obtained from the logarithmic derivative, but requires convergence to decreasing $\Delta \lambda$ at smaller times when the direct derivative is used. We note that due to the symmetry of Eq.~\eqref{eq:Zsym}, it is in principle only necessary to evaluate $\ln Z\left(\lambda,t\right)$ at one value of $\lambda$ to evaluate both the first and second cumulants. In practice we used $\lambda=0$ and $\lambda=0.6$ to eliminate some of the systematic errors due to the finite time discretization.

\section{Results in the presence of interactions\label{sec:Results_int}}

We continue to discuss quantities of physical interest, as obtained from our calculations. In what follows, we will show noninteracting results for comparison, which could be compared to the PI-NEGF data as in the previous section. Such comparisons were performed and agreement was found, but for the sake of brevity this is no longer shown from here onwards. Furthermore, we do not perform the full (and costly) error analysis needed to rigorously define confidence intervals in the inchworm algorithm \cite{Cohen2015,Antipov2017}, but roughly estimate that numerical errors are on the order of several percent.

\subsection{FCS and Shot noise}
The FCS in the presence of interactions is shown at two voltages in Fig.~\ref{fig:ZContour_interacting}, at equilibrium and in the presence of a bias voltage. At first glance, this looks only subtly different from the noninteracting results in Figs.~\ref{fig:V0U0_contour}) and \ref{fig:U0V5_contour}). However, these seemingly small changes in the contour plot encode entirely different physics. To see this, we will explore some of the properties that can be derived from the FCS.

\begin{figure}
\includegraphics{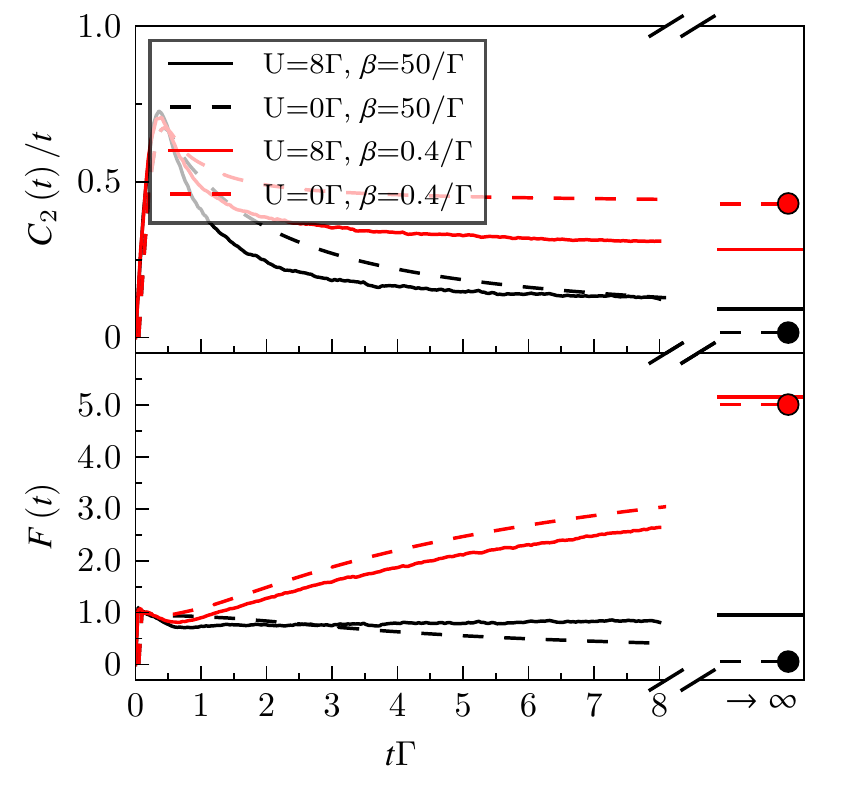}\caption{\label{fig:C1}Evolution of $C_{2}\left(t\right)$ (upper panel) and the Fano factor $F\left(t\right)=C_{2}\left(t\right)/C_{1}\left(t\right)$ (lower panel) for voltage $V=1\Gamma$ and initial condition $\left|0\right\rangle $. To the right extreme of the plot, we show the asymptotic values of these quantities, obtained from linearly fitting the cumulants at long times. The exact values of the noninteracting noise and Fano factor from Levitov--Lesovik theory are shown as filled circles.}
\end{figure}

First, we study the effect of interactions and temperature on the noise and Fano factor. Fig.~\ref{fig:C1} shows the evolution of the time-normalized second moment $C_{2}\left(t\right)/t$ in the upper panel. In the lower panel, the population Fano factor $F\left(t\right)$ defined in Eq.~\eqref{eq:Fanot} is shown, at fixed voltage $V=1\Gamma$ and for the initially empty dot state. The high (low) temperature plots are shown in red (black), and the interacting (noninteracting) results are distinguished by solid (dashed) lines. To the right extreme of the plot, the asymptotic values of the noise and Fano factor are shown. These were obtained from the coefficient of a linear fit to the first and second cumulants $C_{1}\left(t\right)$ and $C_{2}\left(t\right)$, in accordance with the definition of the steady state current and noise defined in Eqs.~\eqref{eq:asympC1} and \eqref{eq:asympC2}. In addition, filled circles mark the Levitov--Lesovik values for the noninteracting quantum noise and Fano factor, obtained from the standard Landauer--B{\"u}ttiker theory \cite{Blanter2000,Ridley2017} via Eqs.~\eqref{eq:levlesexpand-1} and \eqref{eq:Zdotproduct}. We note that the linear fits (which are not shown here) give reliable values to within ~1\% in this case, and in the noninteracting case are perfectly consistent with the Levitov--Lesovik values. The quality of the fits improves with the length of time simulated. In addition we note that asymptotic values shown to the right of Fig.~\ref{fig:C1} are independent of the initial condition.

At high temperatures (red lines), the absolute magnitude of the noise, and also the Fano factor, is enhanced with respect to the low temperature (black lines) case. This is due to the contribution to the second cumulant of thermal or Nyquist noise, which vanishes at small values of $k_{B}T$ \cite{Blanter2000,Ridley2017}. We note that at high temperatures there is an asymptotic divergence between the solid and dashed red curves in the upper panel, as the noise appears to be suppressed by the presence of interactions. 
In the low temperature case, the interacting shot noise is enhanced with respect to the noninteracting value, thus we observe a crossover from noise suppression to enhancement as the temperature is reduced. At low temperature the quantum noise is mainly attributed to the presence of a finite bias \cite{Khlus1987,Blanter2000}. While the thermal-to-shot noise crossover is well described by scattering theory in the noninteracting case, it is far more complex in the presence of strong interactions. 

The low-temperature data corroborates well-known theoretical results by Lesovik \cite{Lesovik1989} and others \cite{Buttiker1990,Blanter1999, Vitushinsky2008}, later confirmed by experiment \cite{Iannaccone1998}, that the presence of inelastic scattering processes in the junction causes excess noise at low temperatures and finite voltages. In particular, inelastic cotunneling processes have been considered, in which electrons tunnel onto and off the dot simultaneously with the creation of a virtual dot state. These leave the dot in an excited state, and have been shown to enhance the noise \cite{Sukhorukov2001,Thielmann2005,Kaasbjerg2015}. When these virtual tunnel states involve correlated electrons of opposite spin, a Kondo singlet is formed \cite{Cronenwett1998}.
At high temperatures, the interaction induces Coulomb blockade on the dot, reducing the number of available transport channels. This suppresses both the current and the noise, such that the Fano factor takes a similar value to that in the noninteracting case (lower panel, red lines). The presence of a large temperature suppresses the formation of coherent states required for inelastic spin-flip and cotunneling processes to occur \cite{Furusaki1995,Konig1998,Tran2008}. However, as the temperature is lowered, we enter the Kondo regime. The Kondo effect increases the inelastic spin-flip rate and facilitates spin fluctuations on the dot, enhancing the noise and resulting in an overall increase in the Fano factor compared to the noninteracting case (solid black lines in Fig.~\ref{fig:C1}). 

\subsection{First passage time distribution}

\begin{figure}

\includegraphics{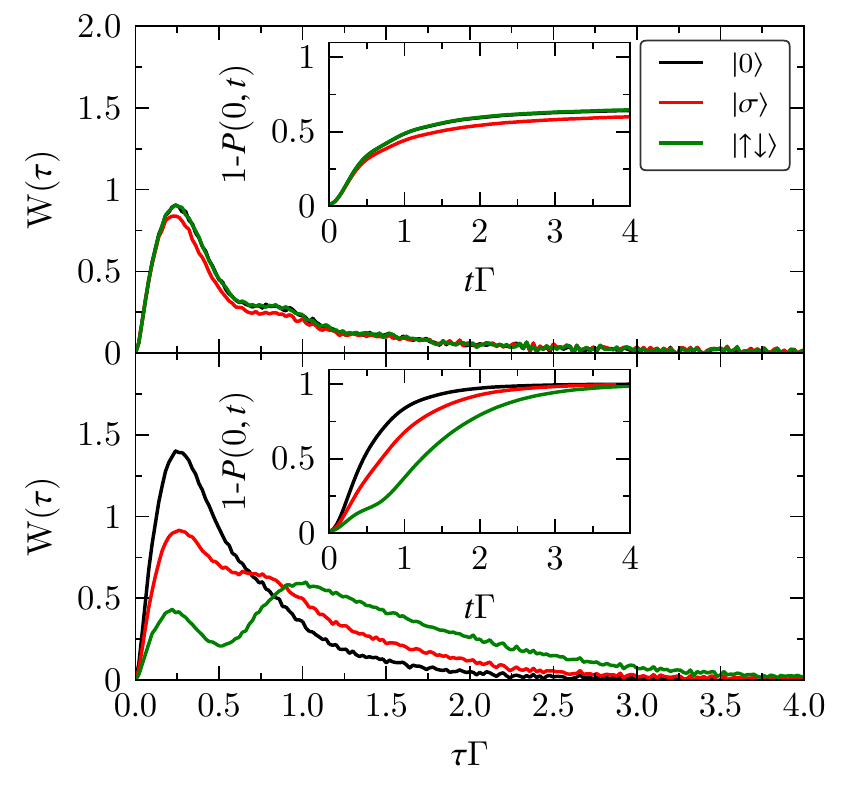}

\caption{\label{fig:WTD_noninteracting}The FPTD $W\left(\tau\right)$ at inverse temperature $\beta=50/\Gamma$ for $U=0\Gamma$, at $V=0$ (upper) and $V=10\Gamma$ (lower). Insets show the integral over the data.}
\end{figure}

\begin{figure}
\includegraphics{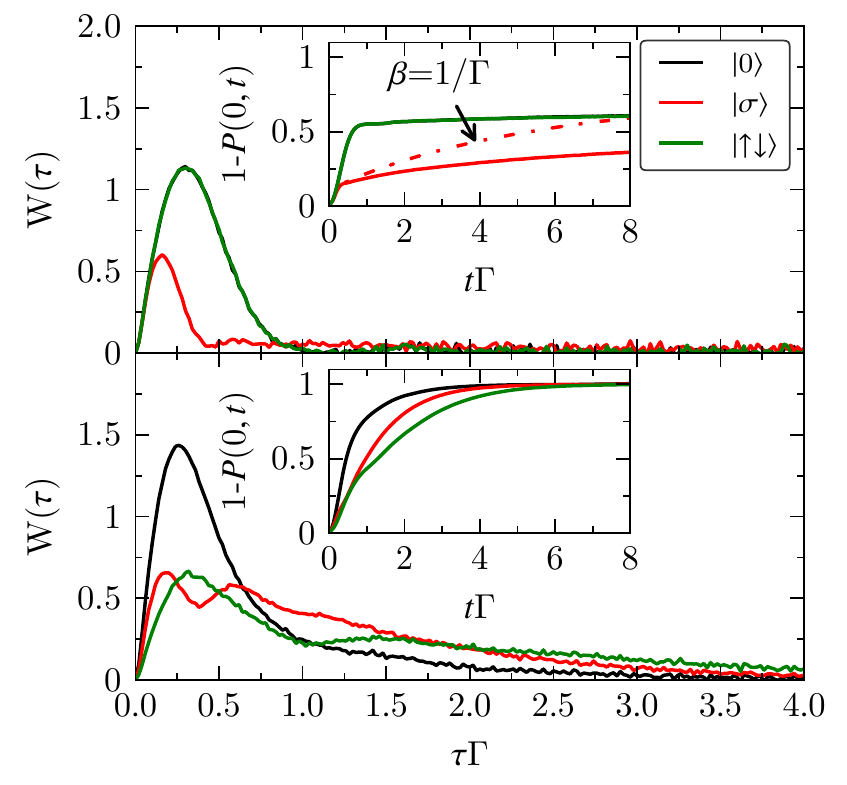}

\caption{\label{fig:WTD_interacting}The FPTD $W\left(\tau\right)$ at inverse temperature $\beta=50/\Gamma$ for $U=8\Gamma$, at $V=0$ (upper) and $V=10\Gamma$ (lower). Insets show the integral over the data, with the upper inset also containing a result at higher temperature (dashed line).}
\end{figure}

In Figs.~\ref{fig:WTD_noninteracting}--\ref{fig:WTD_interacting} we plot the FPTD, as defined in Eq.~\eqref{eq:WTD}, in the unbiased ($V=0$) and biased ($V=10\Gamma$) cases. The inverse temperature in all plots is $\beta=50/\Gamma$, and the dynamics for four different initial states of the dot are shown. Note that the initially half-occupied states $\left|\sigma\right\rangle $ are collected into the same line (shown in red), as the particle--hole symmetric parameters ensure their physical equivalence. Within our data, the two half-filled initial conditions are indeed identical to within numerical errors (not shown).

We begin with the noninteracting problem, Fig.~\ref{fig:WTD_noninteracting}. Here, a maximum occurs at $\Gamma t$ of order $\sim1$, corresponding to the most probable first passage time $\tau_{fp}$, \emph{i.e.} the time at which it is most likely to measure the first change in the left lead's occupation. The first passage probability decays to zero at long times, as the likelihood that the first particle transfer has been detected at very long times becomes vanishingly small. In the unbiased case (top panel), the effect of initial condition on the first passage time distributions is minimal. In particular, the unoccupied (black) and fully-occupied (green) initial conditions are identical in the presence of particle--hole symmetry. This is not the case for the biased system (bottom panel), where electrons are driven out of the left lead. The bias voltage breaks the symmetry between the left and right leads. Therefore---since we are examining the FPTD of the left lead---the symmetry between the doubly occupied and unoccupied initial states is also broken. As the initial occupation of the dot increases in going from the $\left|0\right\rangle $ to the $\left|\sigma\right\rangle $ and $\left|\uparrow\downarrow\right\rangle $ states, the probability density shifts to longer times. A second peak appears in the distribution, corresponding to the first tunneling event on the left lead occurring after an electron has traveled from the dot onto the right lead. This demonstrates an initial condition dependent ``queuing'' effect in the FPTD, which is reminiscent of classical queuing \cite{Lindley1952}.

The insets of Fig.~\ref{fig:WTD_noninteracting} display the integral over $W\left(\tau\right)$, which by Eq.~\eqref{eq:idletime} is equal to the probability $1-P\left(0,t\right)$ that when the total charge of the left lead is measured at time $t$, it has changed from its initial value. In an unbiased system the electron flow across the terminals, as well as the probability that the electron count in the left lead has changed from its initial value, is due entirely to thermal fluctuations and the delocalization dynamics of the wavefunction, and may saturate at a value between one and zero (see upper inset). As the bias $V$ is increased, however, the active driving causes this probability to reach unity rapidly (as in the the lower inset).

Fig.~\ref{fig:WTD_interacting} is identical to Fig.~\ref{fig:WTD_noninteracting}, except for the presence of a finite Coulomb repulsion energy $U=8\Gamma$. At very short times, when the time dependence is linear, the symmetrized interaction has essentially no effect on the FPTD. However, this changes dramatically at timescales substantially smaller than $1/\Gamma$. Comparing the interacting unbiased case, shown in the upper panel, with the corresponding noninteracting result in Fig.~\ref{fig:WTD_noninteracting}, it is clear that the magnitude of the resonant peak and the value of $\tau_{fp}$ are increased for the doubly occupied and unoccupied initial states $\left|0\right\rangle $ and $\left|\uparrow\downarrow\right\rangle $, whereas both quantities are decreased for the half occupied initial states $\left|\sigma\right\rangle $. This can be understood in terms of Coulomb blockade physics: at short times, involving the transfer of the first electron, no difference between the initial conditions is observed. However, in the $\left|\sigma\right\rangle $ case, the suppression of the even charge states by the interaction causes the transfer of a second electron in the same direction to be energetically unfavorable. The opposite occurs in the $\left|0\right\rangle $ and $\left|\uparrow\downarrow\right\rangle $ states.

Another striking difference between the interacting case and its noninteracting counterpart (top panels of Fig.~\ref{fig:WTD_interacting} and Fig.~\ref{fig:WTD_noninteracting}, respectively) is the long tail appearing in the FPTD of the initially magnetized states $\left|\sigma\right\rangle $. The effect of this tail is prevalent in the inset, where very slow relaxation to unity is observed (we show times up to $t=8/\Gamma$ to emphasize the slow relaxation). This can be attributed to the stabilization of local moments \cite{Hewson1997} and slow spin dynamics \cite{Weiss2012} associated with the Kondo regime, to which the system is equilibrating at these parameters. For comparison, the red dashed line in the upper inset of Fig.~\ref{fig:WTD_interacting} shows dynamics for the $\left|\sigma\right\rangle $ initial states at $\beta=1/\Gamma$, where the Kondo effect is suppressed by temperature; the relaxation is then faster.

In the lower panel of Fig.~\ref{fig:WTD_interacting} we apply a bias voltage of $V=10\Gamma$ to the interacting system. At such a large voltage, electrons are chiefly injected into the dot from the left lead and ejected into the right lead, simplifying the analysis of the effects of interaction on the queuing. In the case of the half occupied initial conditions $\left|\sigma\right\rangle $, the first peak is suppressed relative to the second in the presence of interactions, because the energy penalties discussed above decrease the likelihood of electron injection from the left lead before the original electron escapes into the right lead. Similarly, the first peak for the doubly occupied initial condition becomes larger than the second, because the process where two electrons are ejected on the right before the first electron is injected on the left is energetically suppressed. Remarkably, the slow spin relaxation related to Kondo physics is substantially reduced in the inset for the $\left|\sigma\right\rangle$ initial states at this high voltage. This is consistent with claims that a small remnant of Kondo physics can survive at high voltages \cite{Wingreen1994,Anders2008,Cohen2014a}, but requires further investigation.

\subsection{$n$-particle probabilities}

The FPTD, related to $P\left(0,t\right)$, sheds light on the dynamical response of the system to all possible particle tunneling events. It does not distinguish between events in which different numbers of particles are transported. However, the full counting statistics contain much more information: it is possible to access $P\left(\Delta n,t\right)$ for every $\Delta n$. It is of interest to search for manifestations of correlated transport in these detailed distributions. We also consider the most probable time at which the particle number changes by $\Delta n$, given by the maximum of the corresponding distribution. We denote this time by $\tau_{mp}^{\left(\Delta n\right)}$.

\begin{figure}
\includegraphics{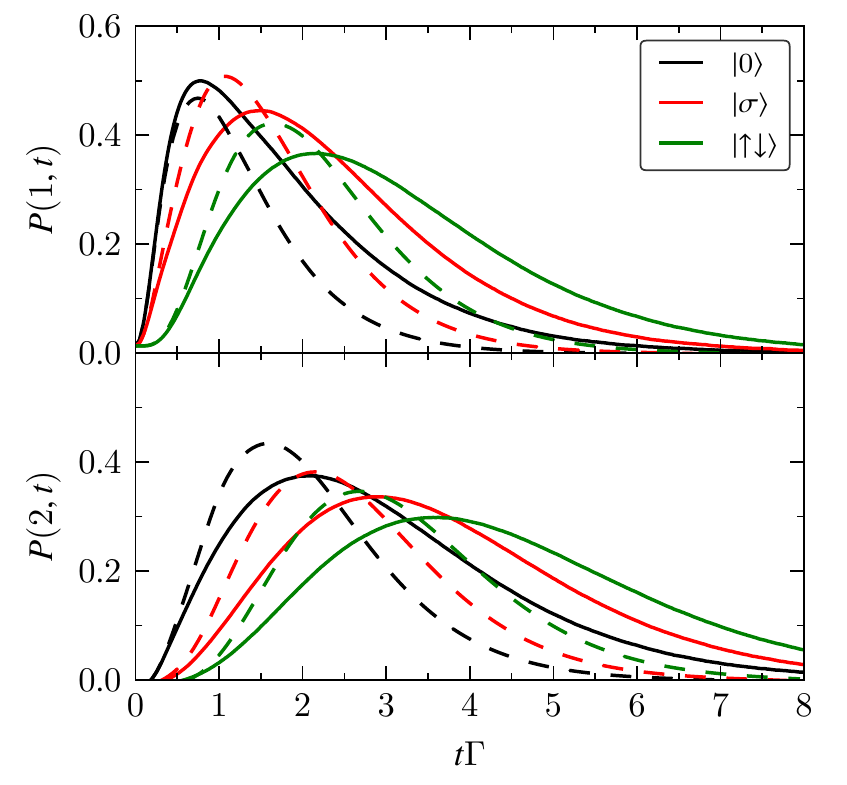}

\caption{The time dependent probabilities for $\Delta n=1$ (upper) and $\Delta n=2$ (lower), for the interacting case $U=8\Gamma$ (solid lines) and noninteracting case $U=0\Gamma$ (dashed lines), with initial conditions shown in the inset. Bias and inverse temperature are set to $V=10\Gamma$ and $\beta=50/\Gamma$, respectively.\label{fig:P_1_t_initial_conds}}
\end{figure}

In Fig.~\ref{fig:P_1_t_initial_conds}, we begin by considering the probability for the particle number in the left lead to have changed by $\Delta n=1$ after time $t$, for each possible initial condition and for the interacting (solid lines) and noninteracting (dashed lines) cases. We once again apply a bias voltage of $V=10\Gamma$, so that it is energetically favorable to move charges from the left lead onto the dot, and charges from the dot to the right lead. Regardless of the presence of interactions, the peak of the probability distributions $P\left(1,t\right)$ for the different initial conditions is shifted to longer times by an increasing number of electrons initially on the dot. In the noninteracting case, this simply reflects the fact that the empty dot has more transport channels open to electron traversal events than a partially or fully filled dot. If the initial state is fully occupied, transport of the first electron is forbidden by the Fermi rule until at least one dot electron has tunneled to (with high probability) the right lead. The effect of Coulomb charging depends on the initial occupation: when the dot begins empty, interaction does not change the position of the maximum, but somewhat enhances the total probability that a single tunneling event was measured at any time, reflecting a reduction in the probability of detecting higher values of $\Delta n$. In this case there is no initial ``queue'' for electron tunneling onto the dot, and the interaction makes the first electron tunneling event more energetically favorable, then suppresses the second. For all other initial states, the Coulomb repulsion shifts the distribution to longer times, as electrons on the left lead cannot enter the dot without ``queuing'' for electrons on the dot to first tunnel to the right lead.

In the lower panel of Fig.~\ref{fig:P_1_t_initial_conds}, we repeat the above analysis for the probability that at time $t$, a change $\Delta n=2$ in the particle number is measured on the left lead. $P\left(2,t\right)$ is shifted to longer times by the interaction for all initial conditions, because the transfer of the second electron must be preceded by that of the first, such that it always encounters some Coulomb repulsion.

\begin{figure}
\includegraphics{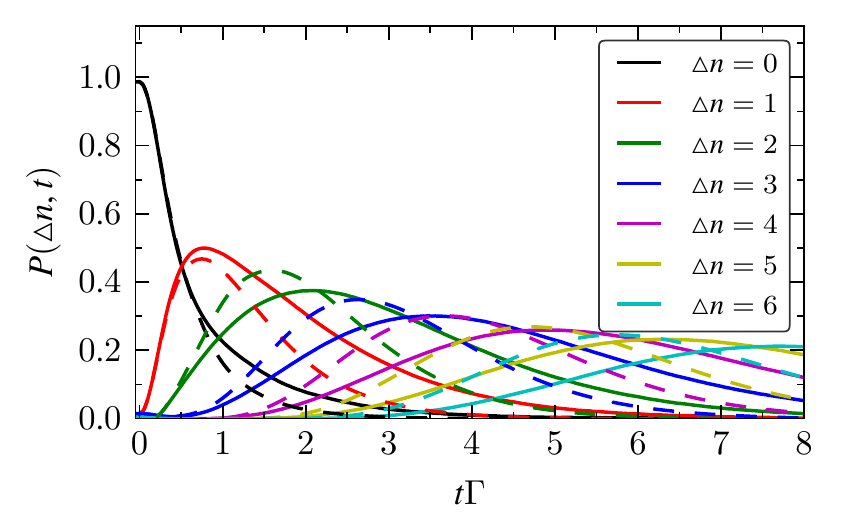}
\includegraphics{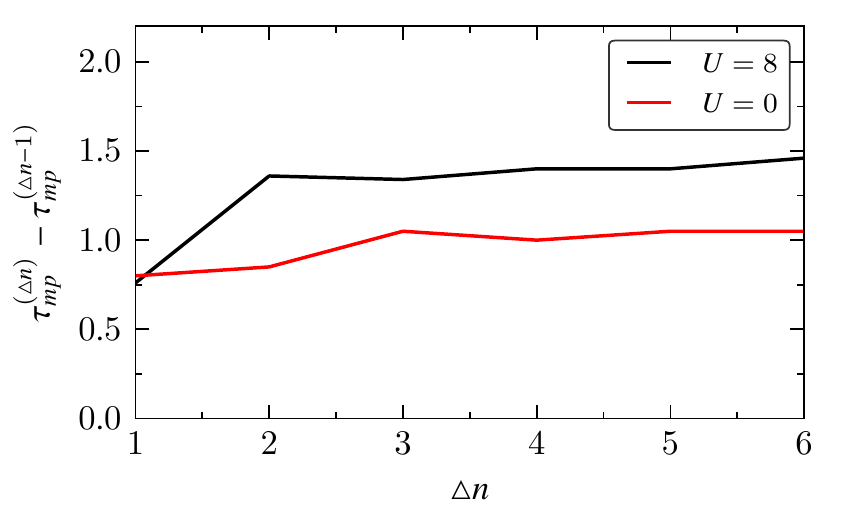}
\caption{\label{fig:P_deltan}Top panel: The probabilities $P\left(\Delta n,t\right)$ with $\Delta n=0,\ldots,6$, for the unoccupied initial state $\left|0\right\rangle $, at $V=10\Gamma$ and $\beta=50/\Gamma$. The solid lines correspond to the interacting case $U=8\Gamma$, and dashed lines correspond to the noninteracting case $U=0$. Bottom panel: The inter-peak distance $\tau_{mp}^{\left(\Delta n\right)}-\tau_{mp}^{\left(\Delta n-1\right)}$ as a function of $\Delta n$, for $U=8\Gamma$ (black) and $U=0\Gamma$ (red).}
\end{figure}

The top panel of Fig.~\ref{fig:P_deltan} shows a sequence of $P\left(\Delta n,t\right)$ for $\Delta n=0,...,6$, focusing on the unoccupied initial state. At short times, we see perfect agreement between the interacting and noninteracting results, as no interaction can take place before some electrons occupy the dot. Increasing the value of $\Delta n$ shifts the $\Delta n>1$ peaks to longer times, which makes sense physically since a change involving $\Delta n=N$ electrons implies all processes leading to a change of $\Delta n=N-1$ electrons have already occurred. We note that we do not show backscattering events $\Delta n<1$, which have a small but finite probability even in the presence of a large voltage. In the unbiased system (not shown), backscattering events are of equal importance to the forward scattering events.

The rich, detailed information on population densities shown in Fig.~\ref{fig:P_deltan} lends itself to a quantitative analysis of electron transfer processes. In the lower panel of the same figure, we study the variation of the distance between maximally probable times $\tau_{mp}^{\left(\Delta n\right)}-\tau_{mp}^{\left(\Delta n-1\right)}$ as a function of $\Delta n$, for the interacting (black curve) and noninteracting (red curve) cases. As we propagate to longer times, this quantity can be considered a proxy for queuing effects at the single electron level at steady state: it describes the typical waiting time between subsequent tunneling events. As $\Delta n$ increases, the peak-to-peak distance stabilizes to a roughly constant value. The long-time value in the noninteracting case is of order $1/\Gamma$, as might be expected for a simple rate process where the only relevant timescale is the coupling between the dot and baths. However, the peak-to-peak distance in the interacting case is significantly larger: as each electron enters the dot, it suppresses the next electron from entering by virtue of the Coulomb repulsion. Interestingly, while the point at which each distribution begins to differ significantly from zero is only weakly modified by the interaction, the width of the distributions and the weight at their tail end are almost immediately enhanced. This may indicate that fluctuations play an increasingly important role in the population transfer dynamics as the strength of the interaction is increased.

\section{Conclusions\label{sec:Conclusions}}

We presented the first numerically exact calculation of full counting statistics for a non-integrable model of interacting fermions, in this case the nonequilibrium Anderson impurity model. Using the inchworm quantum Monte Carlo method, we obtained the generating function $Z(\lambda,t)$ at a variety of physical parameters ranging from the noninteracting case to the strongly correlated Kondo regime. This provides access to currents, which have been accessed before by numerically exact methods; but also to other experimentally measurable quantities which were not. This includes the current noise at steady state, the Fano factor, all higher moments and cumulants of population transfer event, and the complete time-dependent probability distributions for $n$-electron transfer events.

The method was applied to a coupling quench, where a dot is suddenly attached to the leads at time zero and allowed to evolve to equilibrium or to a nonequilibrium steady state. After performing benchmark comparisons in the noninteracting limit to verify the accuracy of our results, we explored the effects of electron--electron interactions and nonequilibrium bias voltage on the full counting statistics and the properties derived thereof. We observed the signatures of the Coulomb blockade and Kondo effects in the noise and Fano factor, and found transient queuing effects depending on the choice of initial condition in the FPTD. By considering the individual, time-dependent probability distributions for the occurrence of $n$ tunneling events, we were further able to argue that such queuing effects persist at steady state. Rather than simply being associated with a lower effective tunneling rate, the rapid widening of the probability densities in the presence of interactions suggests that fluctuations play a significant role in the interacting dynamics.

With the availability of a reliable scheme for calculating the full counting statistics of generic interacting impurity models, a variety of important research questions can now be addressed. Some immediately relevant examples include quantum thermodynamic topics, such as the verification of quantum fluctuation--dissipation theorems and the calculation of efficiency fluctuations in quantum devices; and the evaluation of noise and Fano factors beyond the Fermi liquid regime. We observed queuing effects, but it would also be interesting to look for bunching effects in the presence of an effectively attractive interaction. As the noninteracting physics enters entirely through the hybridization function, it is also feasible to consider the counting statistics of impurities embedded in more realistic noninteracting models of materials and nanosystems, one interesting example being magnetic impurities in graphene nanoribbons \cite{darocha2015,Ridley2017b}.

Spin-dependent FCS in quantum junctions has only recently been treated in the noninteracting case \cite{Tang2017spin}, and spintronic applications are anticipated. The method we have presented can easily be generalized to access spin-dependent counting statistics and multi-lead moments: one could then go beyond the analysis of Fig.~\ref{fig:P_deltan} to consider the time-dependent probability that the left lead lost (\emph{e.g.}) one spin up electron, while simultaneously the right lead gained two spin down electrons. In this context, the long-term survival of spin fluctuations at timescales where the mean magnetization has died out would embody a remarkably clear signature of strong correlation physics. Generalization to include bosonic degrees of freedom is also possible, enabling treatment of thermoelectric systems or coupling to an optical continuum. The method could further be extended to frequency dependent power spectra and related conductance spectra \cite{Rothstein2009,Gabdank2011}, which can be used to study photon absorption and emission processes \cite{Engel2004,Entin-Wohlman2007,Orth2012,Miwa2017}. Additionally, noise in periodically-driven systems with interactions could now be investigated, enabling performance tuning of correlated nanoelectronic devices \cite{Strass2005}. 

To summarize, the full counting statistics of strongly correlated nonequilibrium quantum impurity problems can now be obtained numerically. This provides unprecedented insight into the stochastic nature of electronic transport at the resolution of single tunneling events, and allows simulation of a variety of experiments for which no systematic theory was previously available.

\paragraph{Acknowledgements }
We are grateful to Michael Galperin for directing our attention to this problem. G.C. acknowledges support by the Israel Science Foundation (Grant No. 1604/16). M.R. was supported by the Raymond and Beverly Sackler Center for Computational Molecular and Materials Science, Tel Aviv University. E.G. was funded by  DOE  ER 46932. This research was supported by grant No.~2016087 from the United States-Israel Binational Science Foundation (BSF).

\bibliographystyle{apsrev4-1}
\bibliography{FCS}

\end{document}